\documentclass[11pt,a4paper]{article}
 
\usepackage[margin=1in]{geometry}
\usepackage{amsmath}
\usepackage{amssymb}
\usepackage{graphicx}
\usepackage{bm}
\usepackage{mathtools}
\usepackage{authblk}
\usepackage{cite}
\usepackage{hyperref}
\usepackage{multibib}
\usepackage{xr}
\newcites{supp}{References}

\title{\Large Perturbation-resilient integer arithmetic using optical skyrmions}
\date{}
\author[1,*]{An Aloysius Wang}
\author[1]{Yifei Ma}
\author[1]{Yunqi Zhang}
\author[1]{Zimo Zhao}
\author[1]{Yuxi Cai}
\author[1]{Xuke Qiu}
\author[2]{Bowei Dong}
\author[1,*]{Chao He}
\affil[1]{\small Department of Engineering Science, University of Oxford, Parks Road, Oxford, OX1 3PJ, UK}
\affil[2]{\small Institute of Microelectronics (IME), Agency for Science, Technology and Research (A\text{*}STAR), 2 Fusionopolis Way, Innovis \#08-02, Singapore 138634, Republic of Singapore}
\affil[*]{\small Corresponding authors: aloysius.wang@gmail.com, chao.he@eng.ox.ac.uk}

\begin{document}
\maketitle
\vspace{-12pt}
{\bf The decline of Moore’s law coupled with the rise of artificial intelligence has recently motivated research into photonic computing as a high-bandwidth, low-power strategy to accelerate digital electronics. However, many modern-day photonic computing strategies are analog, making them susceptible to noise and intrinsically difficult to scale. Optical skyrmions offer a route to overcoming these limitations through digitization in the form of a discrete topological number that can be assigned to the analog optical field. Apart from an intrinsic robustness against perturbations, optical skyrmions represent a new medium that has yet to be fully exploited for photonic computing, namely spatially varying polarization. Here, we propose and experimentally demonstrate a method for performing perturbation-resilient integer arithmetic with optical skyrmions and passive optical components. To the best of our knowledge, this is the first time such discrete mathematical operations have been directly achieved using optical skyrmions without external energy input.}\\

Recent developments in structured light have enabled the generation of optical skyrmions \cite{Tsesses2018, He2022, Sugic2021, Shen2022, Bai2020, Lin2021, Cisowski2023, he2023universal, du_deep-subwavelength_2019, lei_photonic_2021, shi_strong_2020, teng_physical_2023, shen_optical_2023, Shen2021_Super, Gao2020, Ma2025, superhighorder, 10.1063/5.0265736}, which include complex spatially varying polarization fields that carry information through a topological number taking values in the integers. Optical skyrmions have three crucial properties that make them ideal candidates for high-density data transfer \cite{shen_optical_2023}, namely the ability to interface with digital information given the discrete nature of the skyrmion number, a robustness to perturbations, and the potential to store arbitrarily large integers within a single localized analog optical field.

These same reasons also make optical skyrmions (here referring specifically to Stokes skyrmions) a natural candidate for computation, with its integer-valued topological number offering a route to digital photonic computing in a way that transcends the usual bitwise framework of digital electronics. Moreover, compared to existing photonic computing strategies, which predominantly modulate amplitude, phase, and wavelength, complex spatially varying polarization fields represent an untapped dimension that can be independently manipulated and, therefore, has the potential to increase bandwidth significantly. This is especially relevant given the growing recognition of untapped spatial degrees of freedom as a means to advance photonic technologies, with free-space optical skyrmions being one such approach \cite{Shen:25, Shen2025Fly}. 

\begin{figure}[!t]
    \centering
    \includegraphics[width=0.95\textwidth]{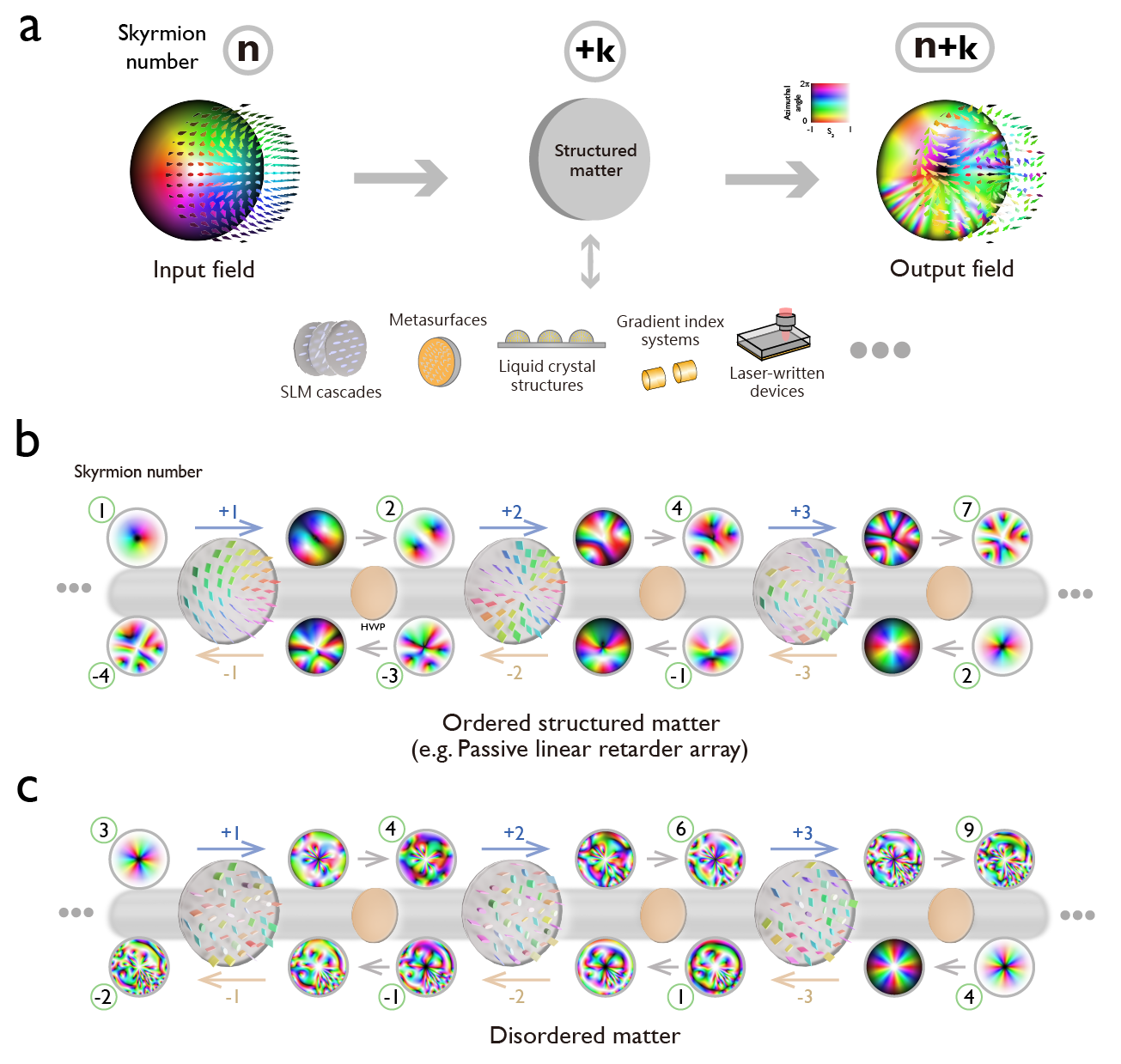}
    \caption{{\bf Concept.} {\bf a,} An optical skyrmion of arbitrary order $n$ passing through specially designed passive structured matter can have the effect of addition or subtraction by an arbitrary integer, $n \mapsto n\pm k$ (only $n \mapsto n+k$ is shown here). Note that the same medium can be used to perform both addition and subtraction. Such structured matter can be realized in many different ways including spatial light modulator (SLM) cascades, metasurfaces, inkjet printing of liquid crystal structures, gradient index systems, direct laser writing of birefringent structures in silica, and more. This figure also depicts the Stokes fields of a standard N\'{e}el-type skyrmion passing through such a medium. Throughout this paper, color is used to represent azimuthal angle on the Poincar\'{e} sphere (PS), and saturation to represent height (similar to \cite{shen_optical_2023}). {\bf b,} Stokes fields of skyrmions passing through different adders of first, second, and third order, and their respective skyrmion numbers. The linear retarder array described in the main text is used as an example, with spatially varying material properties illustrated using cylinders, where the local axis orientation determines the shape and color, and local retardance determines the height. Half-wave plates (HWPs) that control addition and subtraction are also shown (see main text for details). {\bf c,} Adders of first, second, and third order using highly disordered materials, demonstrating the robustness of our proposed adder to imperfections of the medium. Here, we consider perturbations to the linear retarder arrays that respect the conditions proposed in the main text, resulting in ellipticity of the axes and changes in retardance. Note that the chosen disorder here is merely an example that abstractly represents an arbitrary level of distortion.}
\label{fig:concept}
\end{figure}

The central reason for exploring the use of optical skyrmions in photonic computing is, however, their resistance to noise and perturbations which arises from the integer-valued nature of the skyrmion number. The topological robustness of skyrmions has already been established in various domains, including magnetic skyrmions \cite{Nagaosa2013, je_direct_2020, cortes-ortuno_thermal_2017}, and more recently, non-local quantum skyrmions \cite{ornelas_non-local_2024, ornelas2024topologicalrejectionnoisenonlocal}, where the idea of using skyrmions to digitize information is developed from a different perspective. In the setting of optical skyrmions carried through polarization fields, the study of topological robustness has also recently garnered attention \cite{Liu2022}, with a degree of protection already verified for propagation through a number of different media \cite{wang2024topological, he2023universal, teng_physical_2023}. This robustness is particularly relevant to photonic computing, which has thus far remained analog and, therefore, susceptible to noise. Take, for instance, the following three common photonic accelerators. The accumulation of random phase errors in Mach-Zehnder interferometer meshes \cite{shen_deep_2017,f_shokraneh_single_2019}, thermal crosstalk in microring resonator weight banks \cite{tait_neuromorphic_2017, a_n_tait_microring_2016} and low optical contrast in phase-change material-based photonic crossbar arrays \cite{feldmann_parallel_2021} all lead to degraded signal-to-noise ratio in large-size implementations, circumventing the scalability of these architectures. Even with progress in device innovation \cite{zhou_-memory_2023, song_machzehnder_2021}, system calibration, and control algorithm optimization to minimize errors\cite{huang_demonstration_2020,komljenovic_-chip_2018}, almost all photonic accelerators demonstrated thus far have remained small in size (typically less than 4 inputs by 4 outputs). While recent breakthroughs have enabled larger-scale architectures \cite{Ahmed2025Universal, Hua2025IntegratedPhotonicAccelerator}, it has been noted that noise remains a key challenge and a limiting factor for high-throughput data manipulation in photonic chips due to the use of analog photonic signals in computing. Lastly, while spatial degrees of freedom have been utilized for computation-related applications through orbital angular momentum before \cite{mingu, Hong2023OpticalComputingVectorConvolution, Zhang2023OAMClassification}, these approaches do not exhibit the same robustness to noise as they are generally linear-algebraic rather than topological in nature. Computing using optical skyrmions is, therefore, a promising solution offering greater scalability through improved stability against noise. 

Motivated by this possibility, we describe here a class of structured matter (see Methods 1) which, when restricted to particular types of input skyrmions, behaves as adders/subtractors (Fig.\ \ref{fig:concept}a) and provide experimental evidence supporting this fact. Note that the same medium can perform both addition and subtraction, but we adopt the term ``adder'' throughout this paper for brevity. We would also like to emphasize that skyrmions are not topologically protected through the types of structured matter described in this paper, which have been specially designed to manipulate the skyrmion number. However, the topological nature of the skyrmion gives rise to resilience against perturbations in a different form, namely that material parameters have the flexibility to fluctuate without affecting the medium's ability to perform arithmetic (see Methods 1). As mentioned above, this resilience of function is of key importance in photonic computing. Such structured matter can be realized using continuously varying retarders with specific structures on their boundary and offer a reliable way of manipulating the skyrmion number without external energy input. Moreover, we demonstrate that by adopting a generalized skyrmion number \cite{wang2024generalizedskyrmions}, it is possible to simultaneously increase the dimensionality of information carried while relaxing boundary restrictions. Our work opens the doors to this entirely new framework for performing photonic computing, where optical skyrmions are used as the fundamental unit in computation. 

\section{Main}
Here, we introduce a family of optical skyrmion adders: one for conventional skyrmions and another for generalized skyrmions. Each class of adder functions differently and exhibits a distinct form of topological robustness, which we make precise below. Finally, we discuss the relative strengths and weaknesses of the different adders introduced.

Beginning with conventional skyrmion adders, we provide a mathematical description of our proposed optical component, followed by experimental results that demonstrate its feasibility and robustness to disorder. As detailed in Methods 1, given a general homotopy of skyrmions, the resulting difference in skyrmion number is equal to the skyrmion number of the homotopy when restricted onto the walls of the homotoping cylinder. This transformation law gives a systematic way of engineering structured matter to perform addition. 

For example, consider a spatially varying retarder, which when restricted to its boundary, is linear and has a retardance of $\pi$. Let $k$ be the number of half-revolutions made by the axis of the retarder traversing counterclockwise along its boundary (Fig.\ \ref{fig:concept}b). Then, with no further restrictions to material properties apart from continuity, one may show
\begin{equation}
    \deg \mathcal{S}' = \left\{\begin{array}{r l}
        \deg \mathcal{S} + k & \text{if $\mathcal{S}$ is right circularly polarized (RCP) on its boundary} \\
        \deg \mathcal{S}-k & \text{if $\mathcal{S}$ is left circularly polarized (LCP) on its boundary}
    \end{array} \right.
\end{equation}
where $\mathcal{S}$ and $\mathcal{S}'$ are the Stokes fields before and after the medium, respectively. We call structured matter satisfying these conditions skyrmion photo-adders of order $k$.  

\begin{figure}[!t]
    \centering
    \includegraphics[width=0.83\textwidth]{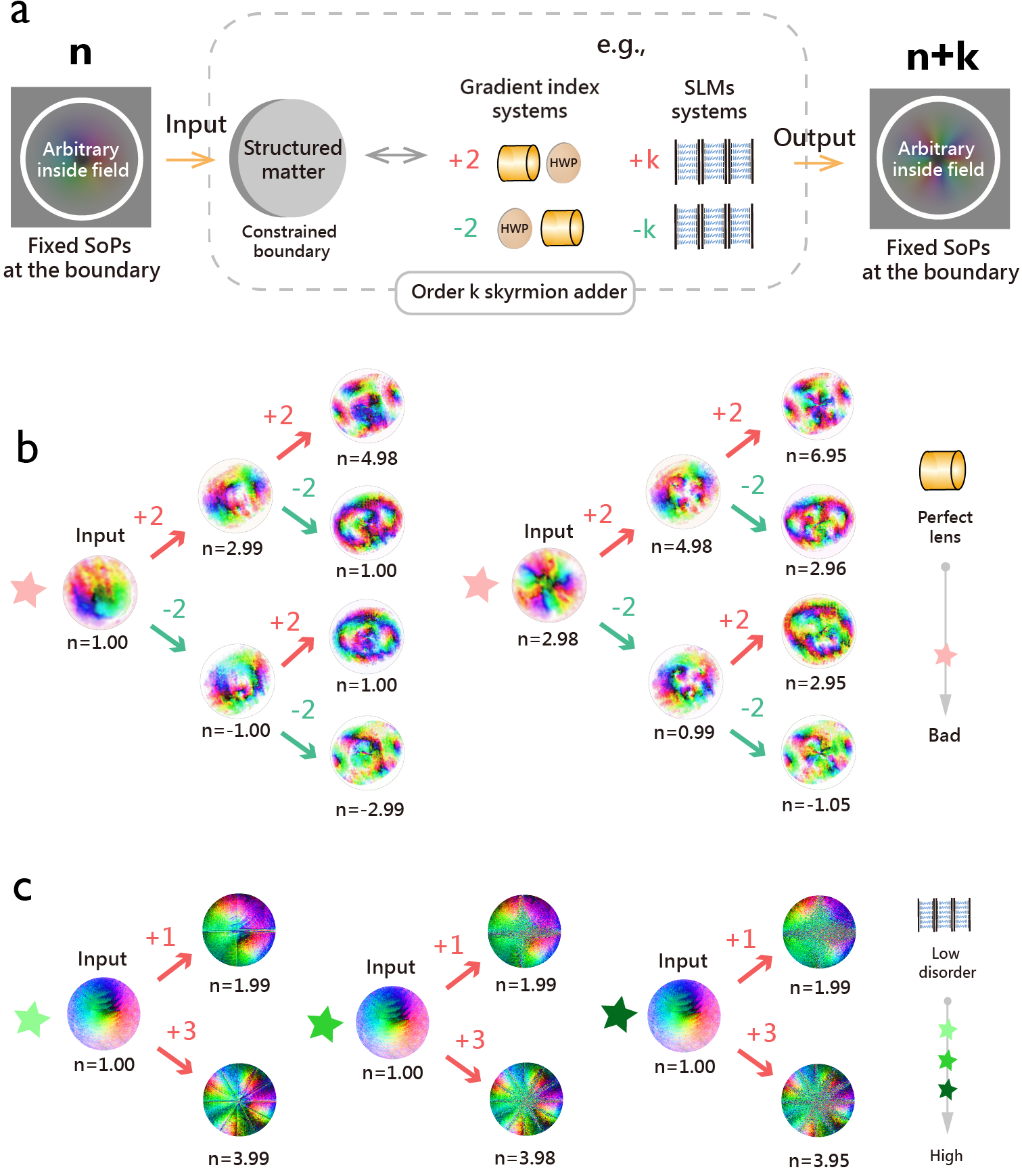}
    \caption{{\bf Adder modules and experimental results.} {\bf a,} Order $k$ adder modules can be constructed using a linear retarder array placed before a half-wave plate while order $k$ subtractor modules can be constructed using a linear retarder array placed after a half-wave plate. The addition of the half-wave plates resolves incompatibilities in polarization states on the boundary as explained in the main text, allowing for the different modules to be cascaded indefinitely. Notice also that order $k$ adders and subtractors can be realized using the same hardware, with one direction performing addition and the other performing subtraction. Lastly, while gradient index systems and a 3-SLM cascade are used to implement the adder in our work, it is worth emphasizing that an adder can be implemented in many different ways, provided its properties at the boundary are properly constrained. {\bf b,} A subset of the measured Stokes fields of optical skyrmions passing through adders of order 2 realized using gradient index systems. The operations $1\pm 2 \pm 2$ and $3 \pm 2 \pm 2$ are shown with detailed implementation presented in Supplementary Note 1 and the full dataset presented in Supplementary Fig.\ 2. {\bf c,} A subset of the measured Stokes fields of optical skyrmions passing through adders of order 1 and 3 realized using a 3 SLM cascade, and where disorder is introduced by a random pixel-wise noise to the voltage levels of the SLMs. Since only a single adder is used in this experiment, no half-wave plate is included. Three levels of disorder are shown as indicated by the color of the star, increasing from left to right. Details of the implementation can be found in Supplementary Note 1, and a complete dataset is presented in Supplementary Fig.\ 3. }
\label{fig: module&experiment}
\end{figure}

Our mathematical results also imply that the function of any medium designed in this way depends only on the structure of its boundary and is independent of material properties everywhere else provided they are continuous (excluding certain extreme cases; see Discussion and Methods 1). This is a reflection of the topological structure of the optical skyrmion exhibited in matter. From a practical perspective, this suggests a strong robustness of any such adder to physical imperfections of the medium that implements it, which greatly eases fabrication. We demonstrate the robustness arising from this ``topological duality'' between field and matter in Fig.\ \ref{fig:concept}c, which shows the invariance of the effect of our proposed adder on the skyrmion number of optical fields to perturbations in the medium, and give a more detailed explanation of robustness and duality in Methods 1. 

After passing through the adder proposed above, the boundary conditions of the skyrmion will flip, that is, RCP becomes LCP and vice versa. Therefore, to cascade multiple adders together, one must realign the outer boundary after each operation. With our proposed design, this amounts to adding a half-wave plate after the medium for addition and a half-wave plate before for subtraction, provided we design our components for skyrmions that are RCP on their boundary (Fig.\ \ref{fig: module&experiment}a and Supplementary Fig.\ 6). Therefore, combining an adder with a half-wave plate effectively produces a cascadable module that performs addition in one direction and subtraction in the other.  

Next, to demonstrate the feasibility of our proposed medium, we provide experimentally measured Stokes fields of skyrmions passing through second-order adders realized via gradient index systems. Fig.\ \ref{fig: module&experiment}b shows a subset of our results, with the full dataset presented in Supplementary Fig.\ 2. We generated skyrmions of orders ranging from $-3$ to 3 using a cascade of 2 SLMs and passed the field through the medium with appropriate waveplates to achieve the operations $+2+2$, $+2-2$, $-2+2$ and $-2-2$ (see Supplementary Note 1 for details of the techniques used in beam generation, measurement and analysis, and Supplementary Fig.\ 1 for the experimental assembly). Note the medium we used was discarded by quality assurance due to its asymmetric axis distribution, and it exhibits significant perturbations in material properties compared to the typical sample. Despite this, the numerically computed skyrmion numbers show that our proposed adder efficiently and reliably performs the desired operations, even with imperfections in the medium. Further technical details of the experiments are presented in Supplementary Note 1.  

\begin{figure}[!h]
    \centering
    \includegraphics[width=0.9\textwidth]{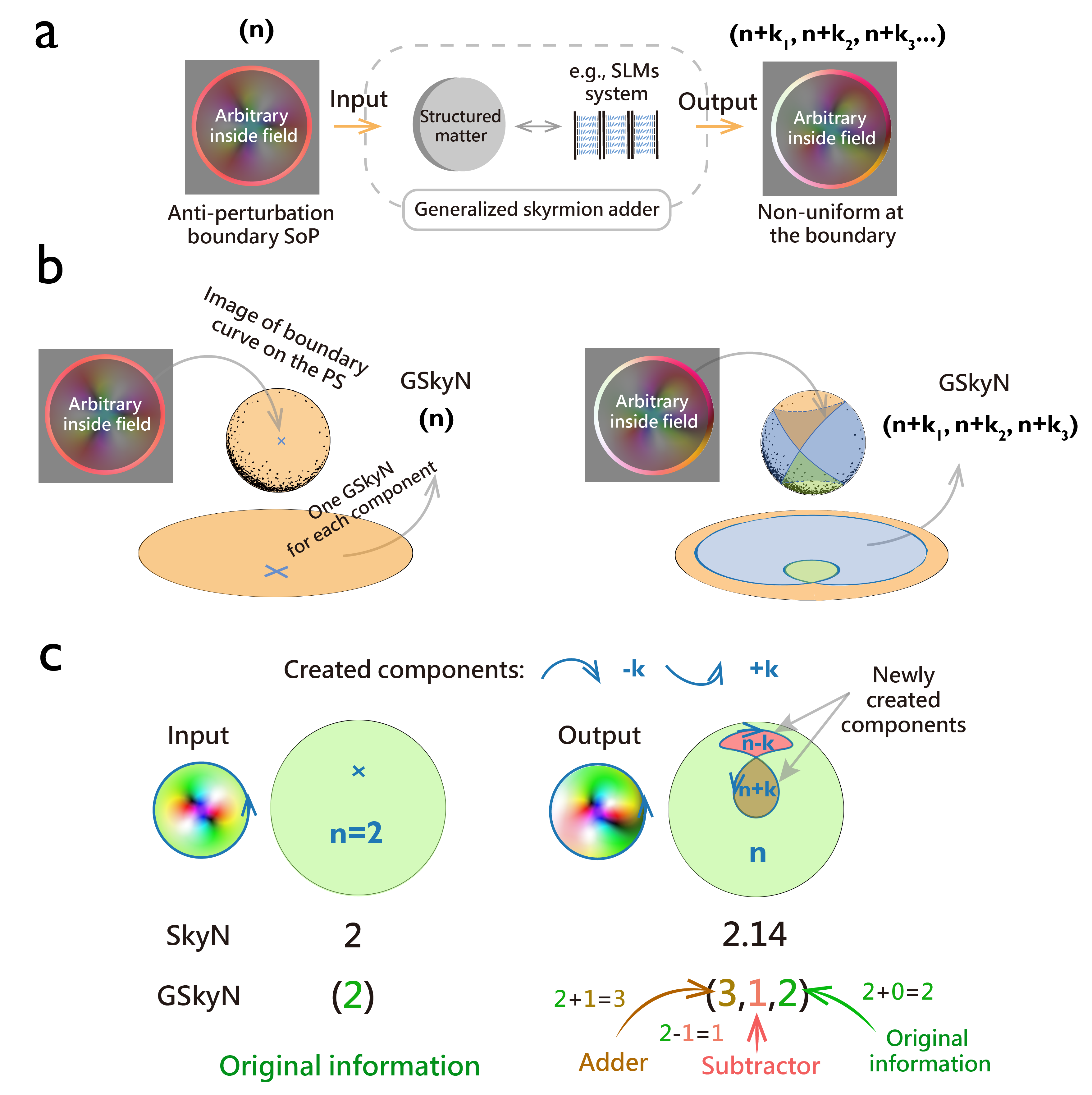}
    \caption{{\bf Generalized skyrmion adders.}  {\bf a,} Concept of a generalized skyrmion photo-adder, which is a passive component that converts a skyrmion of degree $n$ into a generalized skyrmion of degree $(n+k_1, n+k_2, n+k_3, \ldots)$. Note that the function of the adder is robust to perturbations in both the input field and material parameters, with this robustness extending even to situations where perturbations occur at the boundary. {\bf b,} Given a polarization field, a single generalized skyrmion number can be defined for each connected component of the Poincar\'{e} sphere carved out by the image of the boundary curve. A field with (left) one component and (right) three components, along with the corresponding images of their boundary curves on the Poincar\'{e} sphere are shown. A stereographically projected version of the boundary curve is also shown. Note that for a given boundary condition, any continuous extension of the boundary to the entire domain will have the same number of generalized skyrmion numbers. {\bf c,} A generalized skyrmion adder works by manipulating the boundary to create new connected components. For each newly created component, the original skyrmion number is increased once for each time the boundary curve encircles the component, accounting for orientation. The figure depicts an example of a $(n) \mapsto (n+1,n-1,n)$ adder, with input field $n=2$ and where the Stokes fields and stereographically projected boundary curves are shown. Finally, the skyrmion number and generalized skyrmion numbers of the two fields are provided.}
\label{fig: generalized}
\end{figure}

To further emphasize the perturbation resilience of our proposed adder and the versatility of its implementation, we present experimental results where a cascade of 3 SLMs is used to realize an adder and disorder is simulated by introducing random pixel-wise noise to the voltage levels of the SLMs. Since the 3 SLM cascade is designed to achieve arbitrary retardance and axis orientation \cite{he2023universal}, adding noise to the voltage levels of the SLMs effectively simulates a disordered array of arbitrary elliptical retarders, which has the added benefit of reflecting a wide variety of real-world perturbations. Note that the tunability of the SLMs is used to introduce disturbances of varying strengths, but it is not central to how the adder operates (that is, the SLM cascade merely mimics a passive device). The noise is added in such a way that it is maximum at the center and gradually decreases to zero at the boundary, consistent with the derivation in Methods 1. Fig.\ \ref{fig: module&experiment}c shows a subset of our results with the full dataset, including Mueller matrices of the disordered media and polarization ellipses, presented in Supplementary Fig.\ 3. Observe from the figure that the disorder we have added is significantly larger than what would typically occur in practice. Additionally, systematic errors due to phase unwrapping lead to lines observed in the output Stokes field, and this can also be considered a form of perturbation. Nonetheless, the numerically computed skyrmion number remains stable, demonstrating the strong topological robustness of our adder. There remains significant scope for further exploration into the limits of topological protection of optical skyrmions in the presence of random noise, including the effects of spatially correlated noise and the limits at which topological protection breaks down (see details in Supplementary Note 3), which we plan to address in future work.

Lastly, by adopting the generalized skyrmion number introduced in \cite{wang2024generalizedskyrmions}, it is possible to simultaneously enhance the topological robustness of our proposed adder against perturbations in the state of polarization (SoP) of the input light at the boundary, as well as against perturbations in the material parameters at the boundary. Moreover, this approach allows a single field to carry multiple topological charges, representing an increase in the dimensionality of the information carried by the field and significantly improving its information density. 

On a more technical level, the generalized skyrmion number is a method of assigning non-compactifiable fields \cite{wang2024topological} a tuple of integers $(n_1\ldots, n_k)$ derived from the De Rham cohomology of compactly supported forms which are topologically protected under a general notion of homotopy \cite{wang2024generalizedskyrmions}. Given any smooth polarization field $\mathcal{S}\colon \Omega \longrightarrow S^2$, we can define one generalized skyrmion number for each connected component of the Poincar\'{e} sphere carved out by the image of the boundary curve $\mathcal{S}\lvert_{\partial \Omega}$ (Fig.\ \ref{fig: generalized}b). The generalized skyrmion number associated with a connected component can be computed by the integral equation 

\begin{equation*}
    \text{generalized skyrmion number} = \frac{1}{c}\int_{\Omega} f(S) S\cdot \left(\frac{\partial S}{\partial x}\times \frac{\partial S}{\partial y}\right) dx dy,
\end{equation*}
where $c = \int_0^{2\pi}\int_0^{\pi} f(\theta, \phi)\sin\theta d\theta d\phi$ and $f$ is any smooth real-valued function supported on that component. 

Following the line of argument in \cite{wang2024generalizedskyrmions}, one has the following: as a skyrmion of degree $n$ propagates within the medium, the image of the boundary curve on the Poincar\'{e} sphere transitions from a point to a curve. The generalized skyrmion numbers of the newly generated connected components will then be $n+k$ where $k$ is the number of times the boundary curve encircles each component accounting for orientation, while the generalized skyrmion number of the original connected component remains unchanged at $n$ (Fig.\ \ref{fig: generalized}c shows a single example of this process, with more examples given in Supplementary Fig.\ 7). Note that the description above also enables the design of arbitrary generalized skyrmion photo-adders which simultaneously perform an arbitrary number of arbitrary additions, $(n)\mapsto (n+k_1,\ldots, n+k_j, n)$ for any number $j \in \mathbb{N}$ and $k_1,\ldots, k_j \in \mathbb{Z}$. We describe this in detail in Methods 2. 

Specializing to our proposed adder (whose axis distribution is as depicted in Fig.\ \ref{fig:concept}b), if the outer retardance is a constant that lies between $0$ and $\pi$, three cases arise depending on the SoP at the boundary of the incident skyrmion. For SoPs close to RCP, one has the transition $(n)\mapsto (n+k,n)$, for SoPs close to LCP, one has the transition $(n)\mapsto (n-k, n)$, and for all other states, one has the transition $(n) \mapsto (n+k, n-k, n)$. Thus, one can essentially select the adder's function by using different input boundary SoPs.

Notice also that by adopting the generalized skyrmion number, not only can we have situations where multiple additions and subtractions occur simultaneously, but there is also a general tolerance to the boundary SoP of the input beam. Indeed, the adder splits the Poincar\'{e} sphere into various regions, and performs a specific type of operation on each region. More generally, as long as the number of new connected components that form during propagation and the orientation in which the boundary curve encircles each component remains the same, the same operations will be observed at the level of the generalized skyrmion number. This latter property makes the generalized skyrmion adder stable against fluctuations in the parameters of the matter, without the strict constraints at the boundary present in the skyrmion adders introduced earlier. Lastly, note that while it is possible to manipulate fields with non-integer skyrmion numbers for computation, the skyrmion number in this case is not a topologically protected quantity, and therefore does not support the same level of robustness as the generalized skyrmion number (see details in Supplementary Note 2).

\begin{figure}[!h]
    \centering
    \includegraphics[width=1\textwidth]{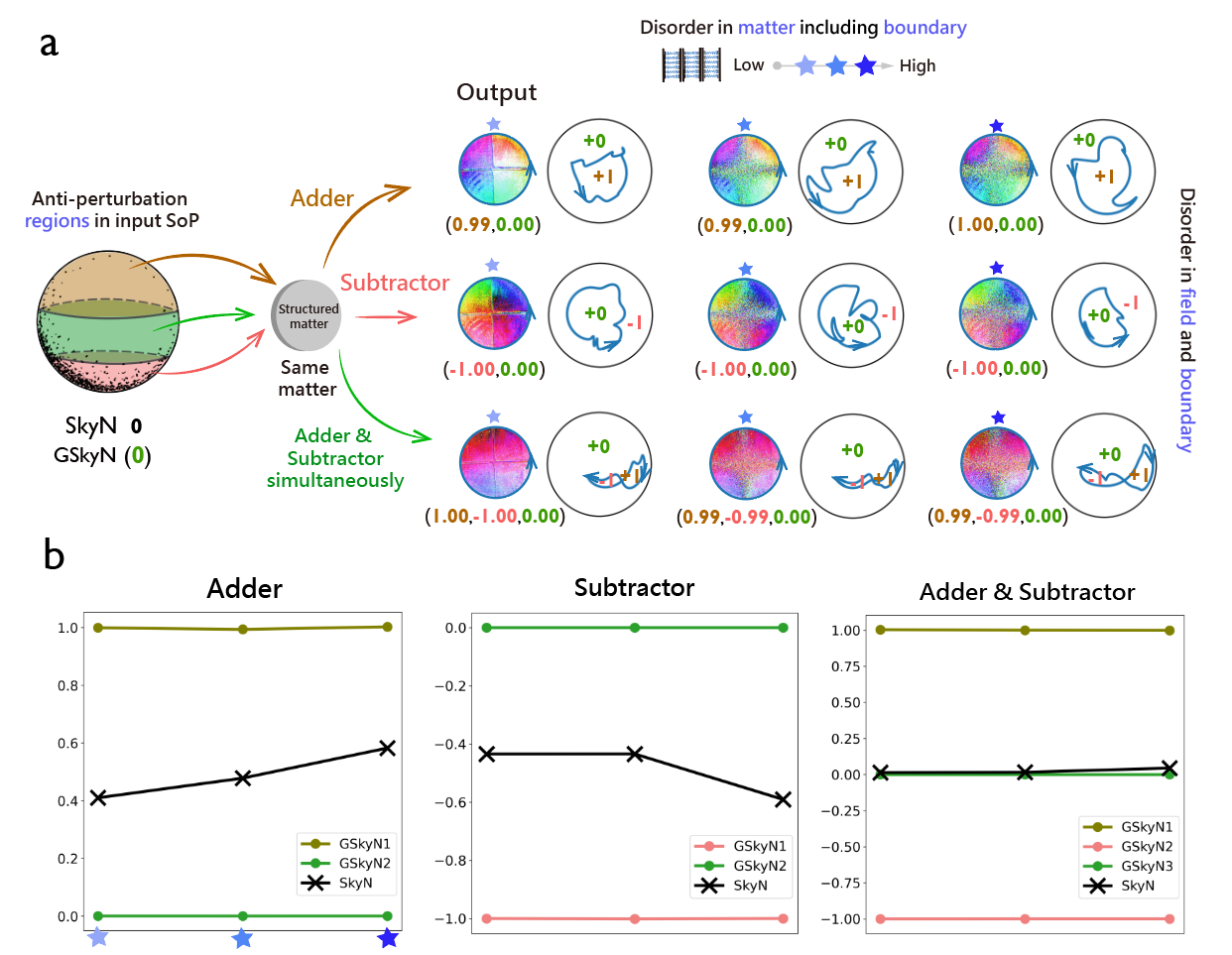}
    \caption{{\bf Experimental results (generalized skyrmion adders).}  {\bf a,} (Left) The different regions of the Poincar\'{e} sphere on which the function of our adder is stable. The top region corresponds to addition, the bottom to subtraction and the middle to both addition and subtraction simultaneously. (Right) A subset of the measured output Stokes fields and computed generalized skyrmion numbers is shown for different levels of disorder (increasing from left to right) and different incident SoPs selected to demonstrate addition, subtraction, and simultaneous addition and subtraction. A stereographic projection of the boundary curve is also shown, with the change in skyrmion number of each region labeled. Note the color of these labels indicate the generalized skyrmion number corresponding to that region. A complete dataset including details of the incident SoPs and measured Mueller matrices is presented in Supplementary Fig.\ 4 . Technical details such as the formation of small loops due to disorder are also discussed. Lastly, details of the implementation and the experimental assembly can be found in Supplementary Note 1 and Supplementary Fig.\ 1, respectively. {\bf b,} The skyrmion number and generalized skyrmion numbers at each level of disorder for different incident SoPs. Note that the generalized skyrmion number is topologically stable even though the usual skyrmion number is not.}
\label{fig: generalized exp}
\end{figure}

To demonstrate the powerful robustness of our generalized skyrmion adder, we present experimental results of uniformly polarized light entering a spatially varying retarder, with axis configuration as in Fig.\ \ref{fig:concept}b with a single half-revolution and where the outer retardance is $\pi/2$. The adder is implemented via a cascade of 3 SLMs as above, and where disorder is simulated by a random pixel-wise noise to the voltage levels of the SLMs, uniform everywhere including at the boundaries. A 2 SLM system which acts as an arbitrary beam generator is separately added (see Supplementary Note 1 for details). Fig.\ \ref{fig: generalized exp}a shows a subset of our results, with a single SoP selected from each region corresponding to addition, subtraction, and simultaneous addition and subtraction. A full dataset including details of the incident light, measured Mueller matrices and polarization ellipses is presented in Supplementary Fig.\ 4. From the computed generalized skyrmion numbers, it is clear that the function of the generalized skyrmion adder is stable with respect to the added disorder. 

Moreover, in the full dataset presented in Supplementary Fig.\ 4, we demonstrate that the function of the adder is stable within a range of SoPs. This stability of the generalized skyrmion adder to different inputs is an important feature which is not enjoyed by the skyrmion adders introduced earlier. Lastly, a comparison between the skyrmion number and generalized skyrmion numbers at different levels of disorders is also shown in Fig.\ \ref{fig: generalized exp}b. From the plots, it is clear that the generalized skyrmion number is topologically stable even though the usual skyrmion number is not. The technique of computing the generalized skyrmion number and estimating the boundary curve is adapted from \cite{wang2024generalizedskyrmions}. 

A more detailed analysis of the experimental results is presented in Supplementary Note 1.4, including discussions on topological protection and engineering solutions for the formation of small loops in the boundary due to disorder. 

In summary, adopting the generalized skyrmion number offers many advantages. Not only are there straightforward methods for creating generalized skyrmion photo-adders capable of performing an arbitrary number of arbitrary integer additions simultaneously, but these systems also exhibit strong robustness against perturbations in both the incident field and the material implementing the adder, with weaker boundary restrictions. 

There are, however, limitations to using the generalized skyrmion number as well, such as difficulties in retrieving the generalized skyrmion number from polarimetric measurements. Our proposed design also only allows for the conversion of a regular skyrmion to a generalized skyrmion, and is therefore not immediately cascadable. In Methods 2, a rudimentary strategy for cascading generalized adders is introduced. Lastly, we note that there remains significant room to explore strategies for directly converting generalized skyrmions to other generalized skyrmions in the context of optical computing. Given its discrete nature, topological robustness and high dimensionality, we believe that the manipulation of generalized skyrmions represents a real and meaningful advancement in the use of optical skyrmions for high-density data applications that extend beyond computing.

\section{Discussion}

In this paper, we have demonstrated a method of achieving digital computing with optical skyrmions using structured matter. In practice, adders of arbitrary orders can be fabricated through numerous techniques including laser polymerization of liquid crystals \cite{zeng_high-resolution_2014, tartan_read_2018}, metasurfaces \cite{yu_broadband_2012, khorasaninejad_broadband_2016, devlin_broadband_2016, wen_helicity_2015, yu_light_2011}, gradient index systems, SLM cascades \cite{he2023universal}, direct laser writing of birefringent structures in silica \cite{fedotov_direct_2016, bricchi_form_2004}, inkjet printing of liquid crystal droplets \cite{kamal_-demand_2023}, compact meta-fibers \cite{he2024metafiber} and more, all of which suggest the possibility of manufacturing compact microscale devices containing such adders.

As mentioned earlier, a crucial property of both proposed skyrmion photo-adders is that its effect on skyrmion fields depends only on the structure of their boundary, a reflection of the topological nature of the optical skyrmion. From an engineering perspective, this provides clear advantages in manufacturing as the tolerable margins of error in fabrication are greatly relaxed, with no further restrictions on the medium besides the continuity of its material properties. In particular, the function of the adder is independent of perturbations to material properties away from the boundary, including spatially varying anisotropic absorptions with tolerance up to a certain diattenuation and complex spatially varying retardance such as those resulting from birefringence (see Methods 1 and \cite{wang2024topological} for more details). Moreover, these conditions can be further relaxed if the framework of the generalized skyrmion number is adopted. 

Despite the numerous promising properties of skyrmions for photonic computing, achieving full skyrmion-based computing still requires solutions to many important problems in manufacturing, architecture, and on-chip integration. Here, we outline possible solutions to some of the hurdles that may arise. 

Firstly, understanding the non-trivial behavior of polarization fields in waveguides is of key importance for enabling small integrated devices to make use of skyrmions. In particular, apart from fabricating the adders themselves, it is also necessary to relay polarization information between adders for any meaningful computation to be done. For a waveguide to support the propagation of complex structured fields, it must necessarily be large enough to carry multiple modes. Moreover, work will need to be done to establish the range of skyrmion numbers that can be achieved by an arbitrary superposition of propagating modes, and the conservation of skyrmion number in propagation. Here, we provide a heuristic argument suggesting the feasibility of topological protection. Based on the ellipticity of the Helmholtz equation, one has, in general, that the electric field develops continuously within the waveguide. This then naturally descends onto a homotopy of Stokes fields provided there are no zeros of the field. Therefore, at least for some finite distance, we expect the skyrmion number to remain unchanged \cite{waveguide}, supporting potential use in optical interconnects, where communication typically occurs over short, chip-scale distances.

Secondly, while our proposed medium theoretically supports wavelength division multiplexing, the generally wavelength-dependent nature of retarders may limit the number of independent operations that can be carried out simultaneously. Nevertheless, with the improving quality of on-chip comb lasers \cite{gaeta_photonic-chip-based_2019} and broadband retarders implemented through metamaterials \cite{yu_broadband_2012, khorasaninejad_broadband_2016, devlin_broadband_2016, wen_helicity_2015}, such limitations are perhaps more a matter of engineering. 

Thirdly, besides considerations relating to the medium, any system exploiting skyrmions will also require auxiliary support for generating and detecting skyrmions. On the front of generation \cite{Lin:24}, notice that uniform polarized light is trivially a skyrmion of degree 0. Therefore, our proposed medium can create optical skyrmions, where incident RCP and LCP light generate skyrmions of degree $\pm k$, respectively. Regarding detection, recent developments in on-chip Stokes polarimetry \cite{zuo_chip-integrated_2023} support the possibility of accurately measuring complex polarization fields on integrated circuits. However, extracting the topological number from polarimetric measurements should be done efficiently for a skyrmion-based photonic adder to be feasible. Note that apart from evaluating the skyrmion number integral, various other properties of the topological number can be used to determine the skyrmion number such as counting strategies involving regular values \cite{wang2024topological}. An alternative way of detection is to exploit light-matter interactions modulated by skyrmions, such as through optomechanical interactions between optical skyrmions and topological solitons in liquid crystals \cite{poy_interaction_2022}. 

Lastly, depending on implementation, both passive and tuneable adders can be fabricated. The best implementation will ultimately depend on the application, with tuneable elements providing greater flexibility at the cost of greater power consumption, possible hysteresis, and greater complexity in control. We reiterate that in our work, we use SLMs to mimic a passive retarder array with disorder, rather than taking advantage of their tunability.

Despite the challenges presented above, we believe that photonic computing using optical skyrmions remains an exciting avenue to explore. Given the rise of artificial intelligence and machine learning, the need for power-efficient computing technologies is more important than ever. Photonic computing has emerged as a promising solution to meet this need, and optical skyrmions represent a way of carrying high-dimensional information within optical fields that holds great potential for enhancing information density in photonic computing without additional energy cost. With wavelength, amplitude, and phase information (including structured phase) accessible independent of spatially varying polarization and, hence, any underlying skyrmion structure, there are certainly intriguing prospects for combining different existing architectures to exploit all these dimensions simultaneously. 

Moreover, our work provides a method of directly exploiting the topological and discrete nature of the skyrmion for computations and, therefore, a route to robust digital photonic computing tolerant to perturbations and noise with strong potential for scalability. This is particularly significant with our introduction of the generalized skyrmion number, which enables the transmission of multiple independent topologically protected quantities within a single field---including those with singularities, which in effect behave as additional boundaries that can be manipulated---allowing for spatial-domain multiplexing in an entirely novel way. We believe this makes our proposed approach one of the most promising strategies for increasing the number of TOPS (trillions ($10^{12}$) of operations per second, or tera-ops per second) in modern photonic processors. Most importantly, using optical skyrmions as units of computation expands the traditional notion of the bit to theoretically infinite values and, therefore, has the potential to alter the binary foundations of digital computing fundamentally.   

As a concluding remark, we note that with addition and subtraction possible, the remaining mathematical operations become a matter of design (see Supplementary Note 4). However, skyrmions also support a more natural notion of multiplication. Suppose a homogenous medium induces a mapping on the Poincar\'{e} sphere of degree $k$. The effect of such a medium on skyrmions is equivalent to multiplication by $k$, with no other restrictions on the input field. For example, a medium whose action on the Poincar\'{e} sphere is given by 
\begin{equation*}
    (\sin\phi\cos\theta, \sin\phi\sin\theta, \cos\phi) \mapsto (\sin\phi\cos k\theta, \sin\phi\sin k \theta, \cos\phi) 
\end{equation*}
is a skyrmion multiplier of order $k$, while a mirror can be regarded as a multiplier of order $-1$ with the proper choice of coordinates \cite{mirror2025}. Lastly, if such a multiplier can be implemented, then division is naturally also possible. One way to achieve this is by taking unique advantage of the fact that skyrmions are integer-valued and not limited to 0 and 1 like bits in conventional digital electronics. This property enables the use of more novel fundamental units of data; for example, a single rational number can be represented by either two conventional skyrmions or a single generalized skyrmion $p/q \Leftrightarrow (p,q)$. Addition, multiplication and division can then be implemented by $(p_1,q_1)+(p_2,q_2) = (p_1q_2+p_2q_1,q_1q_2)$, $(p_1,q_1)\times (p_2,q_2) = (p_1p_2,q_1q_2)$ and $(p_1,q_1)\div(p_2,q_2) = (p_1q_2,p_2q_1)$, respectively. 

\clearpage
\section*{Acknowledgements}
The authors would like to acknowledge the support of the Department of Engineering Science, University of Oxford, and the Royal Society (URF/R1/241734) (C.H.).

\section*{Author Contributions}
A.A.W.\ and C.H.\ conceived the main ideas. A.A.W., Y.F.M., Y.Q.Z., and Z.Z.\ performed the experiments. A.A.W.\ analyzed the experimental results. A.A.W., and C.H.\ prepared the figures. A.A.W., C.H., and B.W.D.\ wrote the paper. A.A.W., Y.F.M., Y.Q.Z., Z.Z., Y.X.C., X.K.Q., B.W.D., and C.H.\ reviewed the results and approved the final version of the manuscript.

\section*{Conflict of Interest}
The authors declare no competing interests.

\clearpage
\bibliographystyle{naturemag}
\bibliography{main}

\clearpage

\section*{Methods}
\setcounter{section}{0}
\section{Structured matter design using a general transformation law of skyrmions}

Here, we derive a general transformation law of skyrmions from which we can observe a powerful heuristic of topological duality, namely that changes to the skyrmion number induced by a medium depend primarily on the structure of the medium on its boundary. As we will explain, this notion of duality has important implications for the design of structured matter to manipulate the skyrmion number, providing a strong tolerance to perturbations of material parameters away from the boundary. 

As an aside, in this paper, we use the term structured matter as an umbrella term for the different types of matter which are used to generate and manipulate structured light \cite{wang2024topological, he2023universal}. Typically, these are complex spatially varying media that can precisely module the phase, polarization and intensity of light, and can be abstractly described using Jones or Mueller calculus. Important classes of structured matter include spatially varying retarders, diattenuators and depolarizers \citesupp{lu_homogeneous_1994}. This framework of structured matter enables the abstract design of complex components, which can then be implemented using different technologies (including metasurfaces, liquid crystal devices, and the various other examples mentioned in this paper) tailored to practical requirements such as size, resolution, tunability, reconfigurability, dynamic control, rewritability (e.g., for information storage), and more \citesupp{ren_reconfigurable_2017, genevet_recent_2017, balthasar_mueller_metasurface_2017}. Additionally, the topological properties of structured matter remain a largely unexplored area with potential applications in beam generation and analysis, biomedical imaging, communications, information storage, and, as demonstrated in this paper, computing.

In the following, we denote by $B_a(0)$ the ball of radius $a$ centered at the origin. Let $\mathcal{S},\mathcal{S}' \colon \overline{B_a(0)} \longrightarrow S^2$ be two different continuous polarization fields that are constant on the boundary. $\mathcal{S}$ and $\mathcal{S}'$ then descend onto the quotient $\overline{B_a(0)}/{\sim} \cong S^2$ obtained by identifying points on $\partial B_a(0)$, and are therefore skyrmions. 

Suppose we can find a continuous homotopy $F\colon \overline{B_a(0)}\times [0,1]\longrightarrow S^2$ from $\mathcal{S}$ to $\mathcal{S}'$. Let $S^+ = \{(s^1, s^2, s^3)\in S^2: s^3 \geq 0\}$ and $S^- = \{(s^1, s^2, s^3)\in S^2: s^3 \leq 0\}$ denote the northern and southern hemispheres, respectively, and define
\begin{align*}
    &\psi_+\colon \overline{B_a(0)} \longrightarrow S^+, &&\psi_+(r\cos\theta, r\sin\theta) = \left(\sqrt{1-(r/a)^2}\cos\theta, \sqrt{1-(r/a)^2}\sin\theta, \sqrt{1-(r/a)^2}\right) \\
    &\psi_-\colon \overline{B_a(0)} \longrightarrow S^-, &&\psi_-(r\cos\theta, r\sin\theta) = \left(\sqrt{1-(r/a)^2}\cos\theta, \sqrt{1-(r/a)^2}\sin\theta, -\sqrt{1-(r/a)^2}\right)
\end{align*}
so that $\psi_+$ is an orientation preserving diffeomorphism onto its image and $\psi_-$ an orientation reversing diffeomorphism onto its image. Now, let $H\colon S^2\times [0,1] \longrightarrow S^2$ be 

\begin{equation*}
    H(s,z) = \left\{\begin{array}{r l}
        F(\psi_+^{-1}(s),z), & s\in \overline{S^+} \\
        F(a\cos\theta, a\sin\theta, z(1+s_3)), & s\in \overline{S^{-}} 
    \end{array}\right.,
\end{equation*}
where $s = (s_1, s_2, s_3) = (\sin\phi\cos\theta, \sin\phi\sin\theta, \cos\phi) \in S^2$. Then $H$ is a continuous by a gluing argument, and by the homotopy invariance of the degree, we have $\deg H(\cdot, 0) = \deg H(\cdot, 1).$

It is clear that $\deg H(\cdot, 0) = \deg \mathcal{S}$ and $\deg H(\cdot, 1) = \deg \mathcal{S}' + \deg \partial F$ where $\partial F \colon \partial B_a(0) \times [0,1]/{\sim} \cong S^2 \longrightarrow S^2$ with the identifications $(x, 0)\sim (y,0)$ and $(x,1)\sim (y,1)$ for all $x,y \in \partial B_a(0)$ and $\partial F([x]) = F(\iota(x))$ 
for the trivial inclusion $\iota\colon \partial B_a(0) \times [0,1] \longrightarrow \overline{B_a(0)}\times [0,1]$. Therefore

\begin{equation*}
    \deg \mathcal{S} = \deg \mathcal{S}' + \deg \partial F.
\end{equation*}
If, further, $\partial F$ is $C^1$, we have the integral equation 
\begin{equation*}
    \deg \mathcal{S} = \deg \mathcal{S}' + \int_{S^2} (\partial F)^\ast \omega
\end{equation*}
for every normalised $\omega \in \Lambda^2(S^2)$.

In particular, this proves that given a general homotopy of skyrmions, the resulting difference in skyrmion number is equal to the skyrmion number of the homotopy when restricted onto the walls of the homotoping cylinder. To make this statement more intuitive, consider the polarization field as it propagates through the medium. Initially, the boundary of the field is a single point on the Poincar\'{e} sphere. However, as the field propagates, its boundary need not remain constant valued, but instead, traces out a curve on the Poincar\'{e} sphere. Lastly this curve collapses back into a point at the output. This situation is therefore exactly analogous to the ``unwrapping'' of the Poincaré sphere from the north pole to the south pole that is usually used to describe the skyrmion (except here we allow for the unwrapping between two arbitrary SoPs), and indeed, the number of times the Poincar\'{e} sphere is unwrapped is exactly the difference in the skyrmion numbers of the input and output fields. 

From the argument above, it is easy to understand why our proposed adder exhibits resilience to perturbations away from the boundary. This resilience arises because the change in skyrmion number depends solely on how the polarization states at the boundary are ``unwrapped,'' which, in turn, depends only on the material parameters at the boundary. In this sense, there is a ``duality'' between light and matter, where specifying the boundary condition of one imposes a corresponding condition on the other. 

Another way to understand topological robustness is through the theory presented in \cite{wang2024topological}, which provides precise conditions for the topological protection of optical skyrmions across a wide range of media, including spatially varying retarders, diattenuators, depolarizers, and cascades of these elements. By considering the Jones or Mueller matrix of a real adder as a composition of an ideal adder and a matrix encoding the non-idealities, the theory in \cite{wang2024topological} can be applied directly to this additional matrix.

With the procedure developed above, we may prove the following. Consider a continuous spatially varying elliptical retarder $J \colon \overline{B_a(0)}\longrightarrow SU(2)$ given by the parametrization 
\begin{equation*}
    J(\alpha, \delta, \Delta) = \begin{pmatrix}
        \cos^2(\alpha)e^{i\Delta/2}+\sin^2(\alpha)e^{-i\Delta/2} & 2i\cos(\alpha)\sin(\alpha)\sin(\Delta/2)e^{-i\delta} \\
        2i\cos(\alpha)\sin(\alpha)\sin(\Delta/2)e^{i\delta} & \sin^2(\alpha)e^{i\Delta/2}+\cos^2(\alpha)e^{-i\Delta/2}
    \end{pmatrix}
\end{equation*}
where $\mathcal{Q} \circ \alpha \colon \overline{B_a(0)}\longrightarrow \mathbb{R}\mathbb{P}^1 = \mathbb{R}/(x\sim x + \pi)$ and $\mathcal{P} \circ \delta \colon \overline{B_a(0)}\longrightarrow S^1 = \mathbb{R}/(x\sim x+2\pi)$ are continuous maps defining a continuously varying fast axis $(\cos(\alpha)e^{-i\delta/2}, \sin(\alpha)e^{i\delta/2})$, $\mathcal{Q}$ and $\mathcal{P}$ the respective quotient maps, and $\mathcal{P} \circ \Delta \colon \overline{B_a(0)} \longrightarrow S^1$ the corresponding continuously varying retardance. Then the output field $\mathcal{S}'$ is given by
\begin{equation*}
    \mathcal{S}'(x) = A^T\text{Spin}\left(J(\alpha(x), \delta(x), \Delta(x))\right)A\mathcal{S}(x) 
\end{equation*}
where $\text{Spin}\colon SU(2) \longrightarrow SO(3)$ is the usual Spin map 
\begin{equation*}
    \begin{pmatrix}
        a+bi & c+di \\ -c+di & a-bi
    \end{pmatrix} \mapsto \begin{pmatrix}
        a^2-b^2-c^2+d^2 & 2ab+2cd & -2ac+2bd \\ -2ab+2cd & a^2-b^2+c^2-d^2 & 2ad+2bc \\ 2ac+2bd & 2bc-2ad & a^2+b^2-c^2-d^2
    \end{pmatrix}
\end{equation*}
and
\begin{equation*}
    A = \begin{pmatrix}
        0 & 1 & 0 \\ 0 & 0 & 1 \\ 1 & 0 & 0
    \end{pmatrix}.
\end{equation*}

Suppose now that $\mathcal{P}\circ \delta\lvert_{\partial B_a(0)} = [0]$ and $\mathcal{P} \circ \Delta\lvert_{\partial B_a(0)} = [\pi]$. If we further restrict to inputs that satisfy $\mathcal{S}\lvert_{\partial B_a(0)} = (0, 0, 1)^T$, then $\mathcal{S}'$ satisfies $\mathcal{S}'\lvert_{\partial B_a(0)} = (0, 0, -1)^T$ and is therefore a skyrmion. Moreover, we may homotope $\mathcal{S}$ to $\mathcal{S}'$  by
\begin{equation*}
    F(x,z) = A^T\text{Spin}\left(J(\alpha(x), \delta(x), \Delta(x) z)\right)A\mathcal{S}(x).
\end{equation*}
One can then directly compute
\begin{equation*}
    \partial F([\theta, z]) = \begin{pmatrix}
        -\sin(2\partial \alpha(\theta))\sin((2n+1)\pi z) \\
        \cos(2 \partial \alpha(\theta))\sin((2n+1)\pi z) \\
        \cos((2n+1)\pi z)
    \end{pmatrix}
\end{equation*}
for some integer $n\in\mathbb{Z}$ and where $\partial \alpha \coloneqq \alpha \lvert_{\partial B_a(0)}$. Taking $\omega \in \Lambda^2(S^2)$ to be the standard volume form on $S^2$ and assuming sufficient smoothness, we have 
\begin{align*}
    \deg \partial F & = \int_{S^2}(\partial F)^\ast \omega \\
     & = \int_{0}^{2\pi}\int_0^1 \frac{1}{4\pi} \partial F \cdot \left( \partial F_\theta \times \partial F_z \right) dz d\theta
    \\ & = \int_{0}^{2\pi}\int_0^1 -\frac{2n+1}{2}\sin((2n+1)\pi z) \frac{d \partial \alpha}{d \theta} dz d\theta \\
    & = -\frac{1}{\pi} \int_0^{2\pi} \frac{d \partial \alpha}{d \theta} d\theta \\
    & = - \deg \mathcal{Q} \circ \partial \alpha
\end{align*}
where $\partial F_\theta$ and $\partial F_z$ are partial derivatives of $\partial F$ with respect to $\theta$ and $z$, respectively. Therefore
\begin{equation*}
    \deg \mathcal{S}' = \deg \mathcal{S} + \deg \mathcal{Q}\circ \partial \alpha
\end{equation*}
depends only on the number of revolutions made by the fast axis along the boundary of the medium. An identical proof can be used in the $\partial \mathcal{S}\lvert_{\partial B_a(0)} = (0, 0, -1)^T$ case to show
\begin{equation*}
    \deg \mathcal{S}' = \deg \mathcal{S} - \deg \mathcal{Q}\circ \partial \alpha.
\end{equation*}

Lastly, note that while other works \citesupp{Chao2019, Shen2023, Qplate} have investigated similar structured linear retarder arrays for skyrmion generation, our study focuses on a broader class of structured matter (including spatially varying elliptical retarder arrays) and demonstrates, for the first time, the topological properties of such media and their ability to modify the topological number of non-uniform input fields. 

\section{Arbitrary generalized skyrmion adders}

Here, we describe steps to design an arbitrary generalized skyrmion adder which performs the operation $(n)\mapsto (n+k_1,\ldots, n+k_j, n)$ for any collection $k_1,\ldots, k_j\in \mathbb{Z}$ using elliptical retarders.

\begin{enumerate}
    \item Pick an incident boundary SoP with Jones vector $J = (J_1, J_2)^T \in \mathbb{C}^2$. 
    \item Design a smooth curve on the Poincar\'{e} sphere which carves out $j$ components and so that the curve encircles the $i$-th component $k_i$ times accounting for orientation.
    \item Lift the curve from Stokes parameters to Jones vectors. 
    \item Suppose we are working with a circular domain $B_a(0)$. Let $(\gamma_1,\gamma_2) \in \mathbb{C}^2$ be the curve obtained in step 3, and set 

    \begin{equation*}
        \Gamma(\theta) = \begin{pmatrix}
            \gamma_1(\theta) & -\bar{\gamma}_2(\theta) \\ \gamma_2(\theta) & \bar{\gamma}_1(\theta)
        \end{pmatrix}\begin{pmatrix}
            J_1 & -\bar{J}_2 \\ J_2 & \bar{J}_1
        \end{pmatrix}^\dagger \in SU(2)
    \end{equation*}

    where ${}^\dagger$ represents conjugate transpose. Notice that $\Gamma(\theta)J = (\gamma_1(\theta), \gamma_2(\theta))$ by construction. Thus, we need only extend $\Gamma$ to all of $B_a(0)$. 
    \item Let $\Phi \colon SU(2) \longrightarrow S^3$ be the canonical diffeomorphism

    \begin{equation*}
        \Phi\begin{pmatrix}
        a+bi & -c+di \\ c+di & a-bi
        \end{pmatrix} = \begin{pmatrix}
            a \\ b \\c \\ d
        \end{pmatrix}.
    \end{equation*}

    We define $\tilde\Gamma \colon B_a(0) \longrightarrow SU(2)$ by 

    \begin{equation*}
        \Phi(\tilde\Gamma(r,\theta)) = \frac{(r/a)\Phi(\Gamma(\theta)) + (1-r/a)p}{\lVert (r/a)\Phi(\Gamma(\theta)) + (1-r/a)p\rVert}
    \end{equation*}

    where $-p$ is any point not in the image of $\Phi \circ \Gamma$. 
\end{enumerate}
The function $\tilde\Gamma$ from the procedure above describes a spatially varying Jones matrix which performs the desired additions. However, it is worth noting that the higher the order of the adder and the greater the number of numbers added simultaneously, the more challenging it becomes to realize these adders in practice. Nonetheless, we believe there is significant potential to further explore such adders and to develop adders capable of converting between generalized skyrmions. 

As a side note, a generalized skyrmion can be converted back into an ordinary skyrmion using a diattenuator with fixed axes and a spatially varying extinction ratio that approaches infinity at its boundary. In this case, the resultant skyrmion number corresponds to the connected component containing the state parallel to the diattenuator's transmissive axis. This then allows for different generalized adders to be cascaded. 

\clearpage

\section*{Data Availability}

All the main data supporting the results of this study are available within the paper, Supplementary Information and Source data. The data that supports the plots within this paper and other findings of this study are available from the corresponding author upon reasonable request. Correspondence and requests for materials should be addressed to A.A.W.\ or C.H.\

\end{document}


\maketitle
\section{Experimental design and results}

In this section, we describe our experiments in more detail, splitting the discussion into two main parts: beam generation and analysis.

\begin{figure}[!h]
    \centering
    \includegraphics[width=0.85\textwidth]{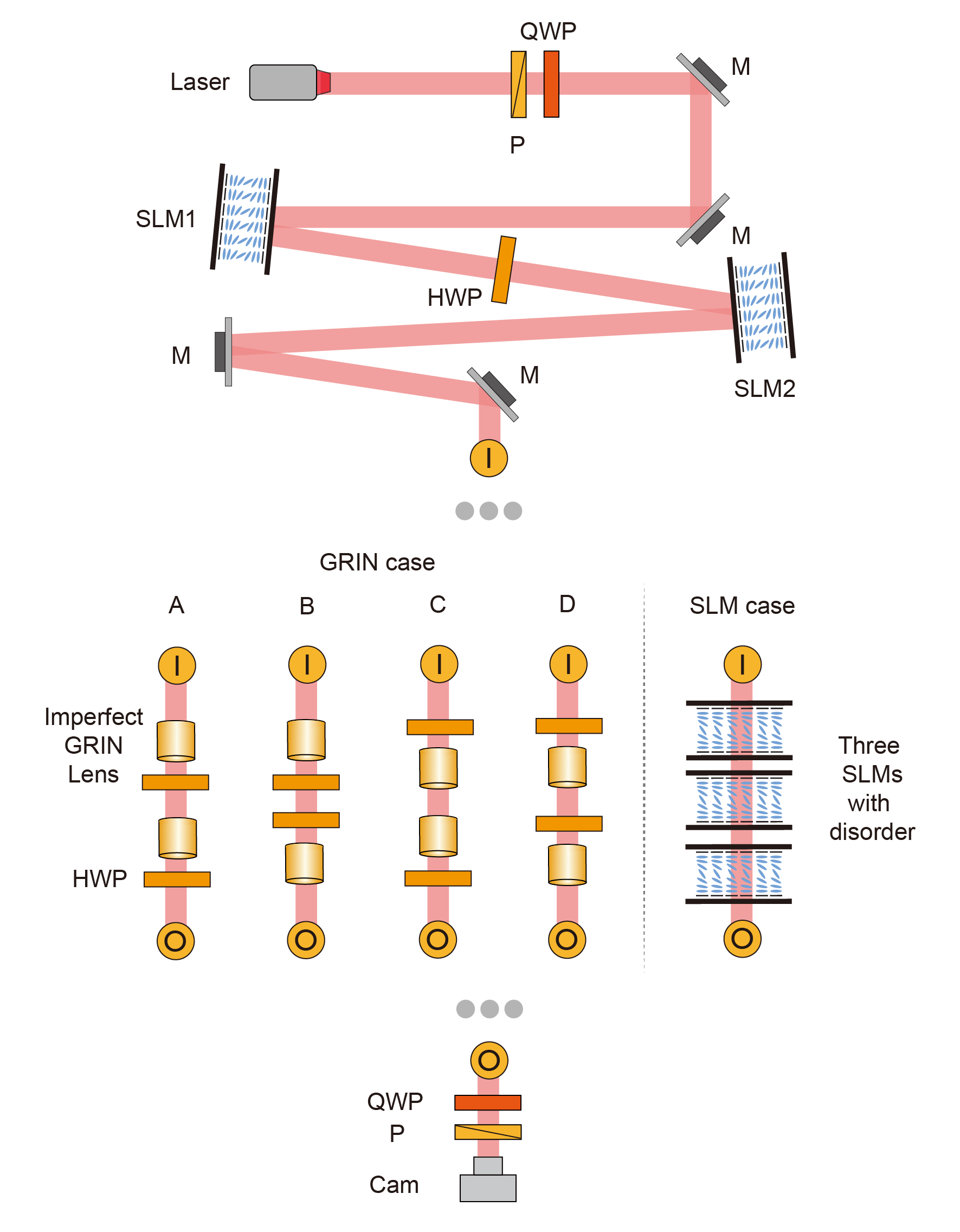}
    \caption{{\bf Experiment assembly.} The complex beam generator and polarimetry configuration adopted in experiments. Components include a He-Ne laser (Melles Griot, 05-LHP-171, 632.8 nm); P: fixed polarizer (Thorlabs, GL10-A); SLM1, SLM2: spatial light modulators (Hamamatsu, X10468-01); HWP: fixed half-wave plate (350-850nm); QWP: rotating quarter-wave plate (Thorlabs, WPQ10M-633); Cam: camera (Thorlabs, DCC3240N). The configurations of the different implementations of the adder used in experiments are also shown. The left (gradient index (GRIN) case) shows the relative placements of modules used to achieve the operations (A) $+2+2$, (B) $+2-2$, (C) $-2+2$, and (D) $-2-2$ while the right (SLM case) shows the 3 SLM cascade used to mimic an adder with disorder.}
    \label{fig:setup}
\end{figure}

The beam generation technique employed here follows from \cite{hu_arbitrary_2020, he2023universal} and consists of two spatial light modulators (SLMs) aligned at a 45\textsuperscript{$\circ$} angle to each other (Supplementary Fig.\ \ref{fig:setup}). The key mathematical insight is that by parametrizing the Poincar\'{e} sphere using spherical coordinates, one has the decomposition

\begin{equation*}
    \begin{pmatrix}
        s_1\\s_2\\s_3
    \end{pmatrix} = \begin{pmatrix}
        \sin\theta \sin \phi \\ \cos \theta \\ \sin\theta\cos\phi
    \end{pmatrix} = \underbrace{\begin{pmatrix}
        \cos \phi & 0 & \sin \phi \\ 0 & 1 & 0 \\ -\sin \phi & 0 & \cos\phi
    \end{pmatrix}\begin{pmatrix}
        1 & 0 & 0 \\ 0 & \cos\theta & -\sin\theta \\ 0 & \sin \theta & \cos \theta
    \end{pmatrix}}_{\text{LC-SLM aligned at 45\textsuperscript{$\circ$} to each other}}\begin{pmatrix}
        0 \\ 1 \\ 0
    \end{pmatrix},
\end{equation*}
so that incident 45\textsuperscript{$\circ$} linearly polarized light on the SLM cascade can be transformed to any other state of polarization (SoP). In our experiments, the incident Stokes field is a standard N\'{e}el-type skyrmion with a beam width of 4mm corresponding to $200 \times 200$ pixels on the SLM. 

To determine the Stokes field, a rotating quarter-wave plate Stokes polarimeter, as described by Azzam \cite{Azzam16}, is adopted. This is one of the most widely used configurations today, and consists of a rotating quarter-wave plate and a fixed horizontal polarizer. The quarter-wave plate rotates at a constant speed, modulating the state of polarization of the measurement channel. The resulting intensity variations are captured by a camera, enabling the calculation of the Stokes vector of the light \cite{Azzam88}. In our experiments, a calibration method based on Fourier analysis \cite{Azzam89} is employed to correct initial azimuth angle errors in the setup, which has been shown to significantly enhance the accuracy of the measured results.

Mueller matrices are measured using a dual-rotating retarder Mueller matrix polarimeter. This system consists of a polarization state generator and a polarization state analyzer \cite{Azzam78}, each formed with a rotating quarter-wave plate and a horizontal polarizer. The state of polarization in the measurement channels is determined by the specific rotation speed ratio of the two quarter-wave plates, which has been optimized to minimize noise propagation during measurement \cite{Smith02}. Calibration of the dual-rotating retarder Mueller matrix polarimeter also relies on Fourier analysis to correct the initial systematic errors in its configuration \cite{Goldstein90}.

\clearpage

\subsection{Experiment assembly}

\clearpage
\subsection{Experimental results (gradient index systems)}

\begin{figure}[!ht]
    \centering
    \includegraphics[width=0.97\textwidth]{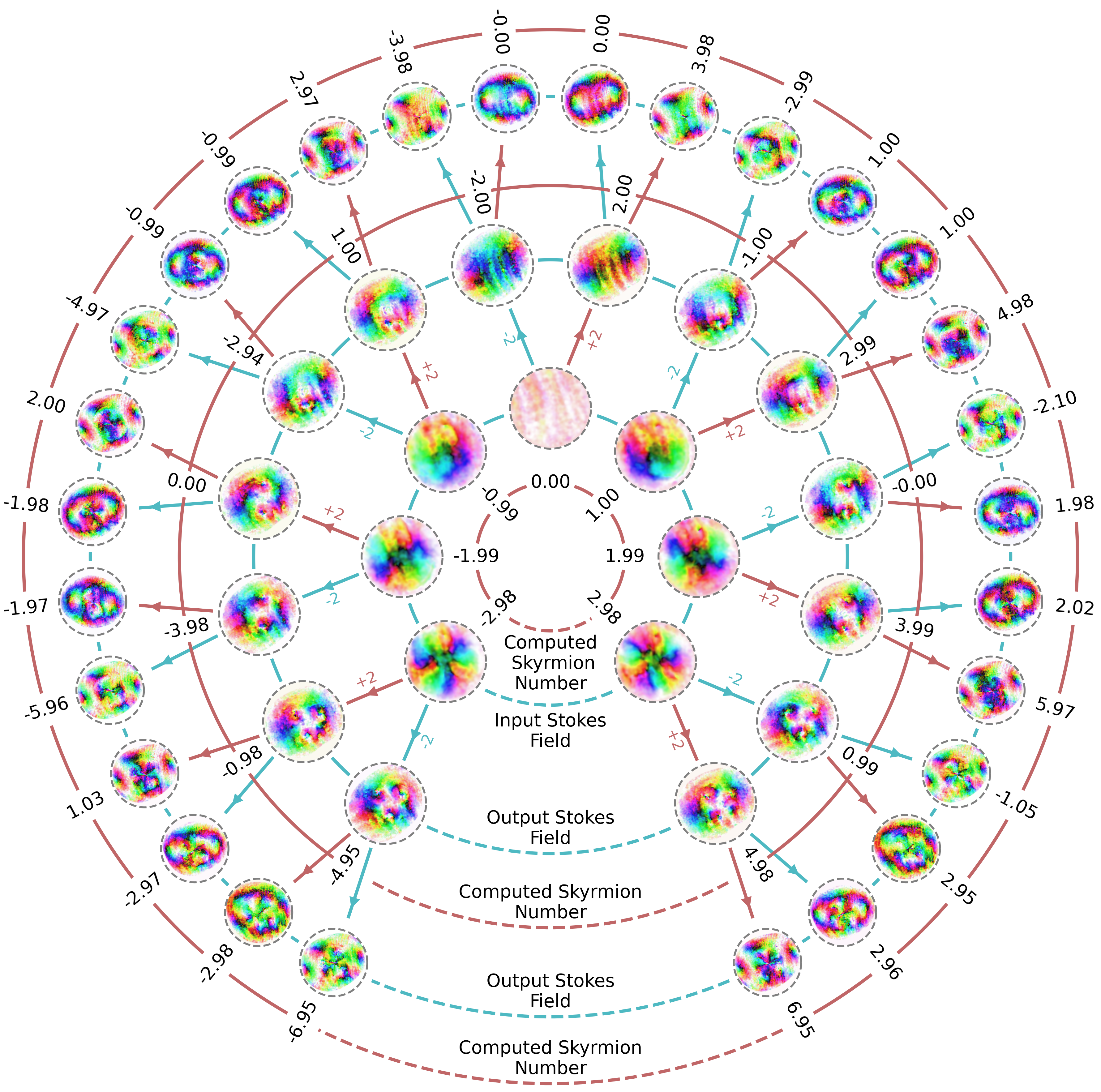}
    \caption{{\bf Experimental results (adders implemented with gradient index systems).} Measured Stokes fields of optical skyrmions passing through adders of order 2 realized using gradient index systems. The innermost ring shows the measured input Stokes fields, the central ring shows the measured fields after a single pass through the medium, and the outermost ring shows the measured fields after a second pass. Here, addition is indicated by red arrows and subtraction by blue ones. Numerically computed skyrmion numbers obtained from measurements are also shown.}
    \label{fig:experiment}
\end{figure}

Supplementary Fig.\ \ref{fig:experiment} shows results for the GRIN lens experiment detailed in the main article. Here, skyrmions of orders ranging from $-3$ to 3 are generated using a cascade of 2 SLMs and passed through the gradient index systems with appropriate waveplates to achieve the operations $+2+2$, $+2-2$, $-2+2$ and $-2-2$. Note from the figure that the numerically computed skyrmion numbers show that our proposed adder efficiently and reliably performs the desired operations. The experiment assembly used is shown in Supplementary Fig.\ \ref{fig:setup}. 

The strategy for computing skyrmion numbers from experimental data is adapted from \cite{wang2024topological}, which employs smoothing to enable accurate computation of partial derivatives required for evaluating the skyrmion number integral. We note that smoothing is a homotopy of the field and therefore does not affect its topological charge. The integration domain is determined based on prior knowledge that the output field should be right-circularly polarized (RCP) at its boundary, following these steps: first, consider the array of data points where $s_3>0.9$. Using these points, fit a curve of the form $r(\theta)$ via Gaussian process regression. The integration region is then defined as the interior of this curve.  

\subsection{Experimental results (disordered adder)}

In this section, we describe experimental results demonstrating the robustness of our proposed adder to disorder. As explained in the main article, a cascade of 3 SLMs is used to realize the adder and disorder is simulated by introducing random pixel-wise noise to the voltage levels of the SLMs. The noise is added in such a way that it is maximum at the center and gradually decreases to zero at the boundary. As mentioned in the main article, systematic errors due to phase unwrapping lead to lines observed in the output Stokes field, and this can also be considered a form of perturbation. Here we demonstrate the operations $0+1$, $0+3$, $1+1$ and $1+3$ at three increasing levels of disorder. The experiment assembly used is shown in Supplementary Fig. \ref{fig:setup}, and the strategy for computing the skyrmion number is adapted from \cite{wang2024topological}.

\begin{figure}[!ht]
    \centering
    \includegraphics[width=0.8\textwidth]{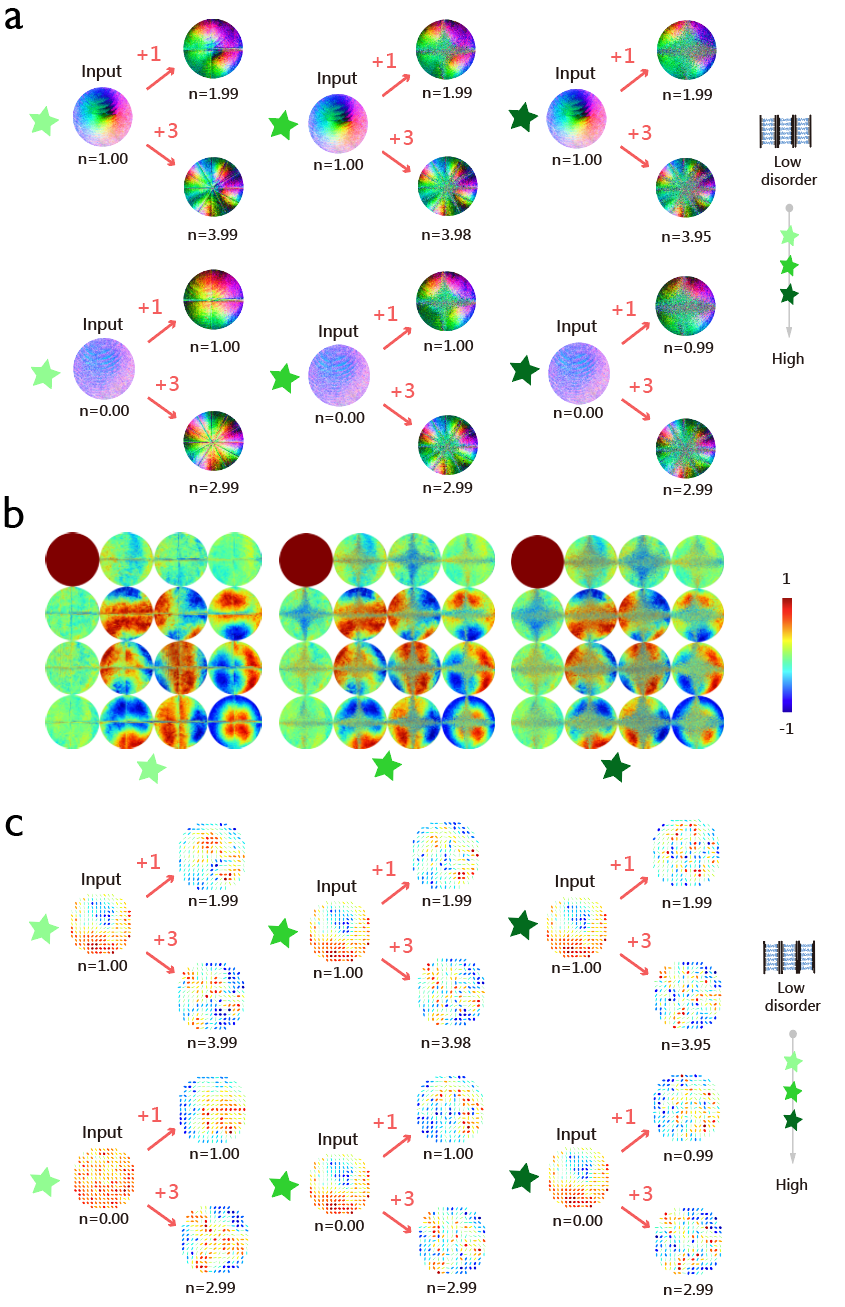}
    \caption{{\bf Experimental results (adders with disorder).} {\bf a,} Measured Stokes fields and skyrmion numbers of optical skyrmions passing through adders of increasing disorder. {\bf b,} Mueller matrices of the implemented adder at different levels of disorder, increasing from left to right. {\bf c,} Experimental results presented using polarization ellipses. }
    \label{fig: adders with disorder}
\end{figure}

Supplementary Fig.\ \ref{fig: adders with disorder} shows the full data, with Stokes fields of the different cases presented both using hue plots and polarization ellipses. The computed skyrmion numbers are also shown. The figure is organized such that the lowest disorder case is presented on the left and with increasing disorder moving from left to right. Measured Mueller matrices of the adder at different levels of disorder are also presented.

\clearpage
\subsection{Experimental results (disordered generalized adder)}

\begin{figure}[!ht]
    \centering
    \includegraphics[width=1\textwidth]{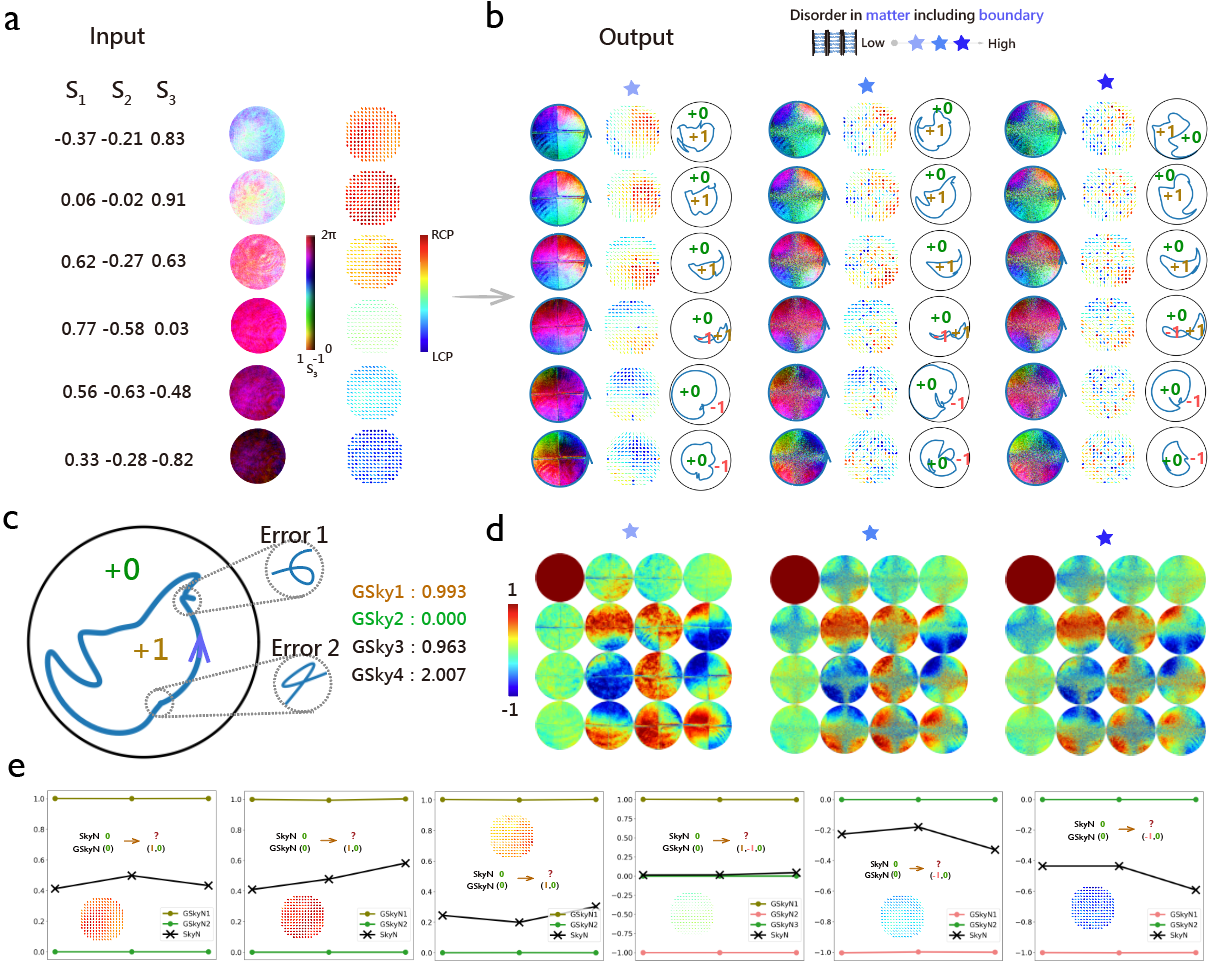}
    \caption{{\bf Experimental results (generalized adders with disorder).} {\bf a,} Measured input Stokes fields presented using both hue plots and polarization ellipses. The average Stokes parameters are also given. {\bf b,} The output Stokes fields along with the computed stereographically projected boundary curves. The orientation of the different curves follows that shown in Fig.\ 4 of the main text. The results are presented with the lowest disorder case on the left, and with increasing disorder moving from left to right. The top three rows demonstrate addition, the fourth row demonstrates addition and subtraction simultaneously, and the remaining rows demonstrate subtraction. {\bf c,} A larger version of the stereographically projected boundary curve corresponding to the second input with medium disorder. Notice that disorder can lead to small loops forming when computing the boundary curve, resulting in additional generalized skyrmion numbers. Here, GSky3 is the generalized skyrmion number associated to the loop causing error 1 while GSky4 is the generalized skyrmion number associated to the loop causing error 2. {\bf d,} Mueller matrices of the implemented generalized adder at different levels of disorder, increasing from left to right. {\bf e,} The computed skyrmion number and generalized skyrmion numbers for each input SoP at different levels of disorder. Here, the position of the figures from left to right correspond to the SoPs in {\bf a} from top to bottom.}
    \label{fig: generalized dders with disorder}
\end{figure}

In this section, we describe experimental results demonstrating the robustness of our proposed generalized adder. Here, we show 6 different input boundary SoPs (Supplementary Fig.\ \ref{fig: generalized dders with disorder}a) passing through adders at three increasing levels of disorder, and whose structure is as described in the main text. The experiment assembly used is shown in Supplementary Fig. \ref{fig:setup}, and the strategy for computing the generalized skyrmion number is adapted from \cite{wang2024generalizedskyrmions}.

The top three datasets in Supplementary Fig.\ \ref{fig: generalized dders with disorder} are different SoPs where addition occurs, while the bottom two are different SoPs where subtraction occurs. As mentioned in the main article, this shows that the function of the adder is stable within a range of SoPs, and there is a general robustness to perturbations of the input. 

Notice also in Supplementary Fig.\ \ref{fig: generalized dders with disorder}c that due to disorder, the computed boundary curve (using the methods of \cite{wang2024generalizedskyrmions}) will sometimes self-intersect and form small loops. While theoretically these loops carry a generalized skyrmion number, they depend on the Gaussian process regression used to estimate the field and are artifacts of the computational strategy. However, notice that despite these loops, the generalized skyrmion number of the large components behave as we expect them to. Thus, it is clear that the larger the size of a component, the greater its topological stability. Following \cite{wang2024generalizedskyrmions}, in the analysis of our data, we use the area of each connected component to determine when to ignore it and treat loops smaller than a threshold size as errors. Note that the non-linear scaling of the stereographic projection map amplifies the apparent areas of loops near LCP; therefore, the sizes of loops shown in Fig.\ 4 and Supplementary Fig.\ 4 do not reflect their true sizes on the Poincar\'{e} sphere. This threshold acts as an engineering metric to overcome small loops that arise from measurements in real-world applications. In the future, different optimization strategies can be designed based on specific application conditions or environmental factors.

\clearpage

\section{Stability of generalized skyrmions vs ordinary skyrmions}

In this paper, we introduced the concept of generalized skyrmions to relax boundary conditions and expand the types of fields that can be used for computations. However, a simpler approach can be taken by using the ordinary skyrmion number for fields with non-integer values. We argue here that the latter approach is less robust. This is because, by construction, the generalized skyrmion number does not take continuous values but rather only values from a discrete set---namely, the set of tuples of integers. This property is what enables the generalized skyrmion number to be stable when the regular skyrmion number is not. That is, since the generalized skyrmion number can only take values in a discrete set, and each value is given by an integral formula (and therefore depends continuously on the field), continuous perturbations of the field cannot change the generalized skyrmion number. Note that this is the same reason why the ordinary skyrmion number is stable provided one can ensure compactification---the boundary condition necessary in ensuring that the skyrmion number integral is integer valued.

However, if compactification does not hold, the skyrmion number becomes unstable as it can now take continuous values, but the generalized skyrmion number is still robust since it remains discrete. For example, consider a typical meron which has polarization field given by 

\begin{equation*}
    S = \begin{pmatrix}
        \sqrt{1-f(r)^2}\cos\theta \\ \sqrt{1-f(r)^2}\sin\theta \\ f(r)
    \end{pmatrix},
\end{equation*}
where $f(r)$ is some function which runs from $f(0)=1$ to $f(1)=0$. Here, the skyrmion number integral evaluates to 0.5. However, notice that if the value of $f(1)$ changes, then so too does the conventional skyrmion number. Therefore, the skyrmion number is not a stable quantity. The generalized skyrmion number, however, remains at $(1,0)$ for all values $-1<f(1)<1$, and is therefore robust to changes in boundary conditions.

Therefore, while it is possible to use non-integer skyrmion numbers for computations, the ordinary skyrmion number in this case is not a topologically robust quantity. It is thus more preferable to use the integer-valued generalized skyrmion numbers instead.

\clearpage

\section{Optical skyrmions in the presence of noise}

The impact of noise on topology is inherently complex, as noise often involves discontinuous functions, whereas topology relies on continuity to be well-defined. Nonetheless, if we are in a setting where noise can be assumed to be continuous, the theory established in \cite{wang2024topological} proves that topological protection of the skyrmion number is guaranteed if the perturbation can be achieved by a compactification-preserving homotopy. For arbitrary spatially varying retarders, the only constraint this imposes is that the retarder is uniform on the boundary. Heuristically, this implies that uniform random noise would alter the skyrmion number (since compactification is disrupted), but topological robustness can persist even if the noise is spatially correlated, provided it remains sufficiently small at the boundaries. 

Note also that we expect the generalized skyrmion number to exhibit greater topological robustness. In particular, it can tolerate noise at the boundary as long as the noise does not destroy the relevant component under consideration. As demonstrated in the main text, our proposed generalized skyrmion photo-adder remains robust even in the presence of uniform random noise. 

For perturbations that are not continuous, the issue becomes more complex. There are two main hurdles that need to be addressed, namely (1) what is largest function space which has a meaningful notion of degree and (2) under what circumstances can we assume that our noisy signal lives in this function space. As it turns out, the degree generalizes to a certain class of distributions, namely functions of vanishing mean oscillation (VMO), and there are Sobolev embeddings $W^{s,p}$ into VMO whenever $sp = n$, $0<s<n$ for functions on $\mathbb{R}^n$. Note, however, that for $n=2$, $W^{1,2}$ embeds into VMO but $L^2$ does not. Since we typically expect uncorrelated random noise to live in $L^2$, working in VMO may not be quite strong enough to establish the theoretical tolerance of the skyrmion number to random noise.

On top of functional analytic considerations, we also need to address the issue of discretization caused by the finite resolution of the camera. In fact, this may be beneficial given the discussion above: although we may not be able to define the degree on $L^2$, in the discretized setting, we can view discretization as the camera detecting an ``area-averaged,'' smoothed-out perturbation. The difficulty, however, lies in making sense of continuity and topological numbers in the discrete context. In this regard, there is significant scope to explore further possibilities and deepen our understanding.

\clearpage

\section{Skyrmion adder architecture}

Here, we describe a possible architecture of an $n$-bit skyrmion adder that is compatible with wavelength division multiplexing. The $4$-bit implementation is shown in Supplementary Fig.\ \ref{fig:architecture}, and can be easily generalized to higher bits by either increasing the number of generator and adder blocks or the number of wavelengths used. The adder works in the following way:

\begin{enumerate}
    \item Suppose we want to perform the operation $a + b$. The number $a$ is first converted from electronic signal to optical signal by selectively activating different sources which pass through the required generator block formed from structured matter as described in our main text.
    \item The output of the generator block is then channelled simultaneously into the adder block as shown in the figure below, which is converted back into electronic signal via a detector array. As we have described the challenges of detection in our main text, we do not reiterate them here.
        
\begin{figure}[!h]
    \centering
    \includegraphics[width=0.85\textwidth]{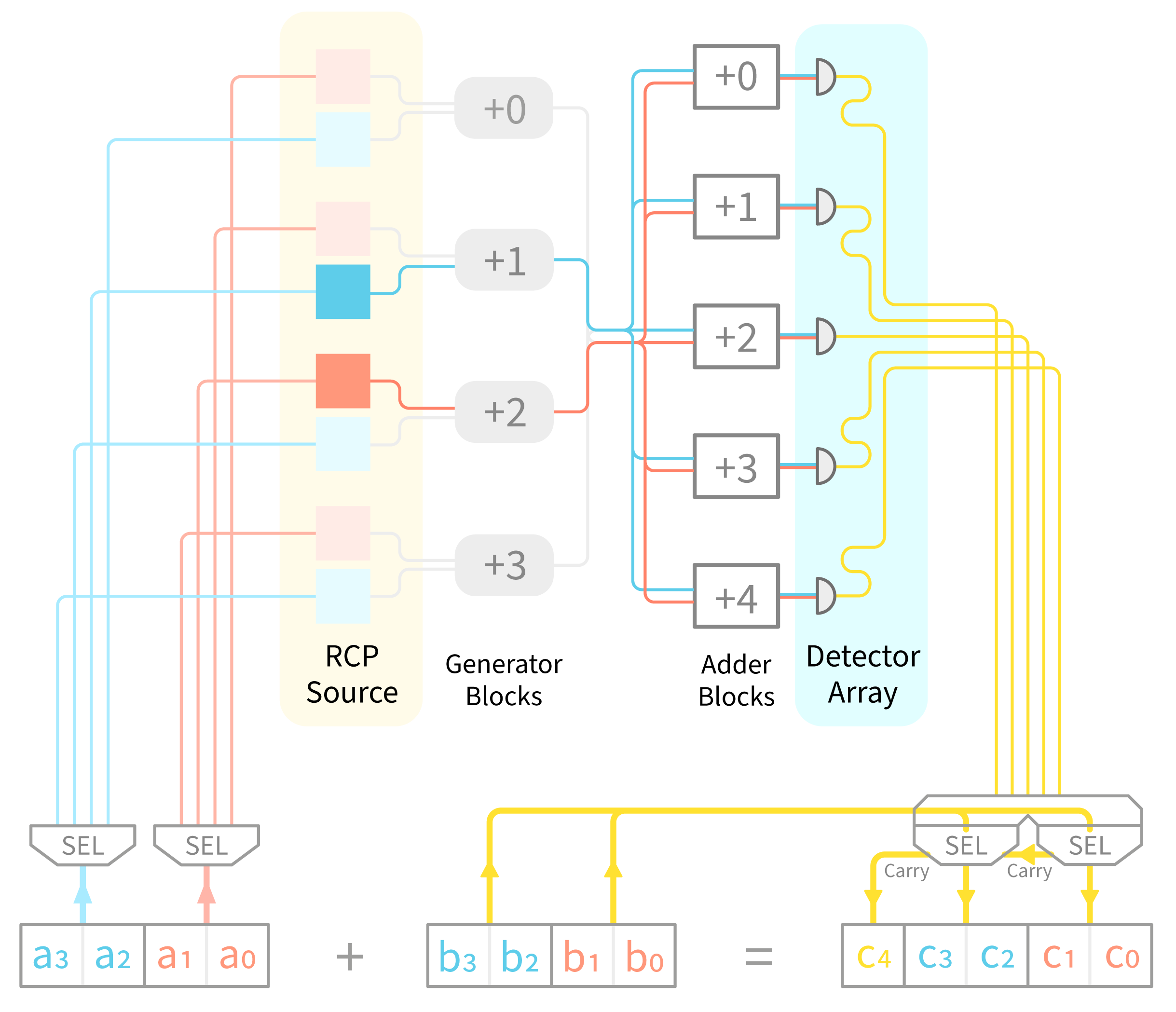}
    \caption{{\bf Possible architecture for an 4-bit skyrmion adder which supports wavelength division multiplexing.} A detailed description of how the adder functions is given in the main text. Here, the gray lines represent waveguides which relay optical signals from one place to another while the yellow lines represent electronic signals. We show two different wavelengths being used at once, represented by the colors red and blue. Note that these colors are intended to distinguish between different independent wavelengths and are not representative of the actual wavelengths of each channel. In practice, comb lasers are a practical approach to generating high-quality beams at different wavelengths, with the wavelengths of the independent channels being closely spaced.}
    \label{fig:architecture}
\end{figure}

    \item The required output from the detector array is then selected using $b$, with carry logic performed, as of right now, using electronic circuitry.
\end{enumerate}
We note that there is still significant scope for improvement in the architecture presented here, particularly with respect to cascadability. This is an important consideration as analog to digital converters (ADCs) and digital to analog converters (DACs) are some of the most power intensive components in photonic chips today. Therefore, performing as many computations in the optical domain as possible before passing to and from electronic signals is one of the most important design considerations. 

Furthermore, as mentioned in the main text, since skyrmions are not constrained by the binary structure of conventional digital electronics, there is potential to explore alternative representations of numbers. For instance, a rational number could be represented by two skyrmions. In this regard, there remains significant room for exploration.

\clearpage

\section{Additional Figures}

\begin{figure}[!h]
    \centering
    \includegraphics[width=0.75\textwidth]{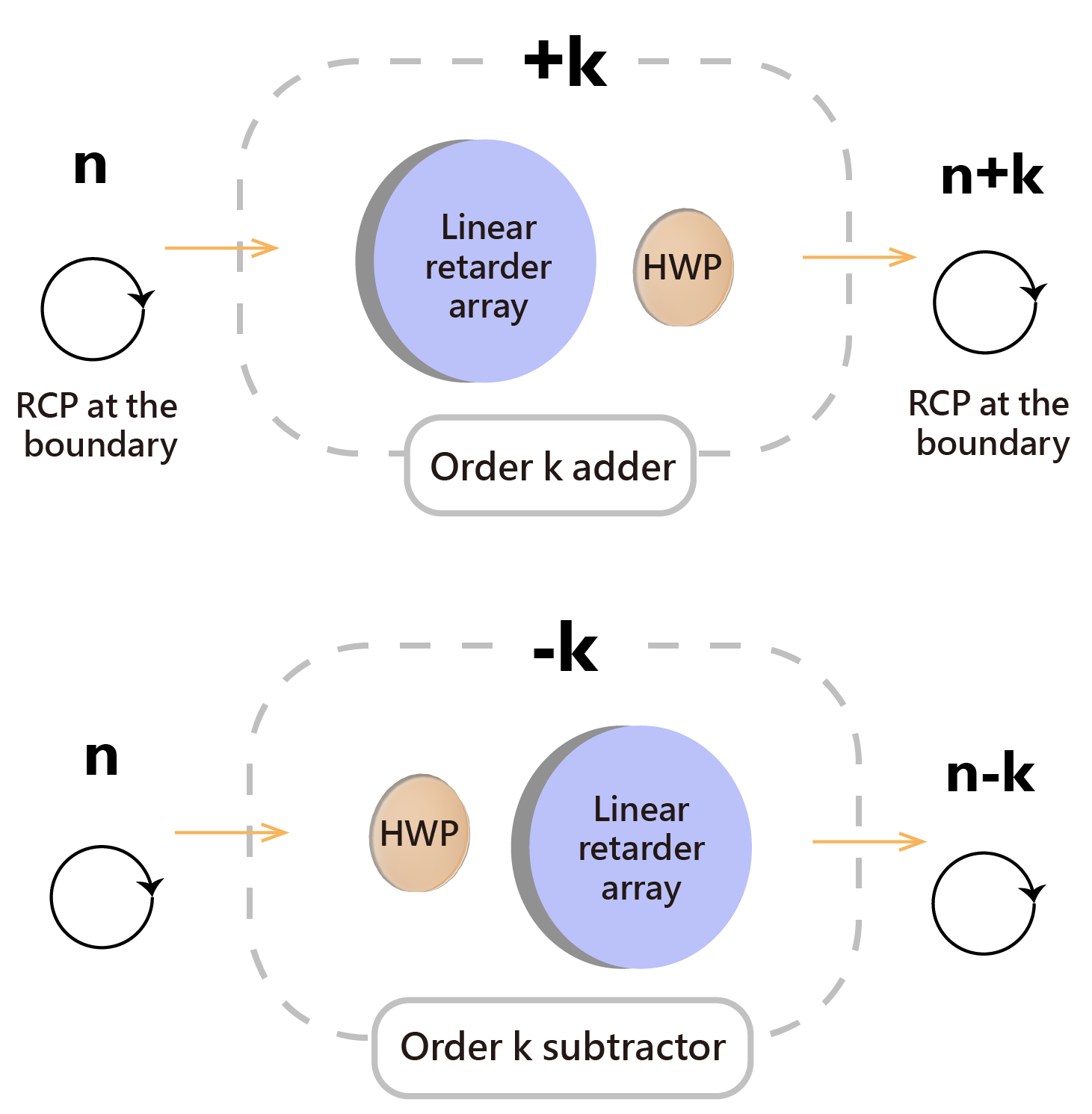}
    \caption{{\bf Adder modules.} The relative placements of half-wave plates (HWP) that enable cascadability of modules designed for inputs that are RCP at the boundary. For addition, the boundary changes from RCP to LCP after passing through the linear retarder array, so a HWP is placed after to restore the boundary back to RCP. For subtraction, the input needs to be LCP at the boundary, so a HWP is place before to adjust the boundary to the correct input polarization state. }
    \label{fig: HWP}
\end{figure}

\clearpage

\begin{figure}[!h]
    \centering
    \includegraphics[width=0.8\textwidth]{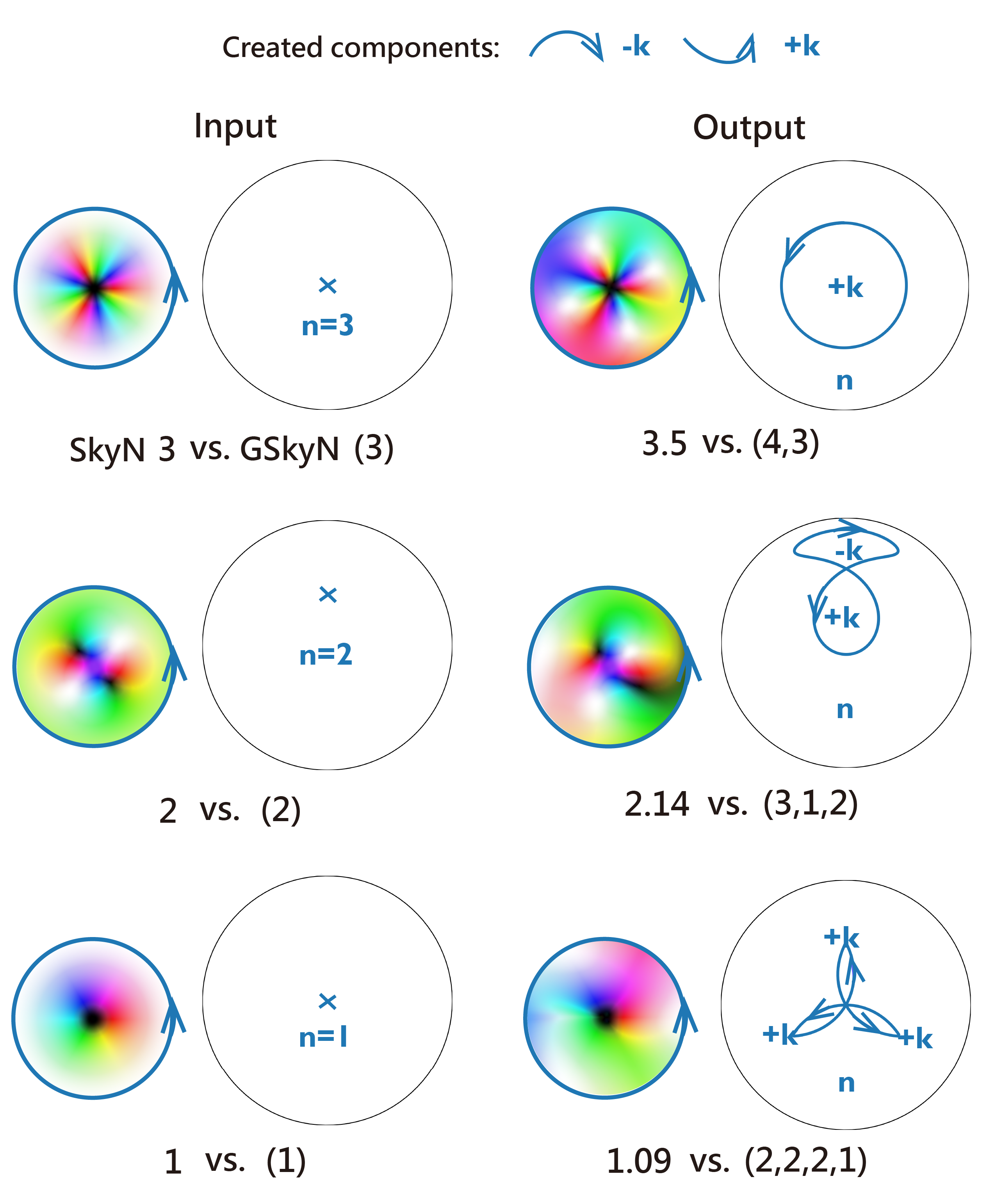}
    \caption{{\bf Generalized skyrmion adders.} As mentioned in the main article, a generalized skyrmion adder works by manipulating the boundary to create new connected components. For each newly created component, the original skyrmion number is increased once for each time the boundary curve encircles the component, accounting for orientation. The figure depicts, from top to bottom, examples of $(n)\mapsto (n+1)$, $(n)\mapsto (n+1,n-1,n)$ and $(n)\mapsto (n+1,n+1,n+1,n)$ adders, where the medium implementing each adder is designed using the strategy presented in Methods 2. Different input and output Stokes fields, along with their stereographically projected boundary curves, are also presented. Finally, the skyrmion number and generalized skyrmion numbers of the various fields are provided. }
    \label{fig: GSkyN-example}
\end{figure}